\documentclass[11pt]{article}
\usepackage{amsfonts,amsmath,amssymb}
\usepackage{enumerate}
\usepackage{hyperref}
\usepackage{bbm}
\usepackage{nicefrac}
\usepackage[all]{xy}
\usepackage{graphicx}
\usepackage{bm}
\usepackage{makecell}
\usepackage[table]{xcolor}

\usepackage{upgreek}

\usepackage{booktabs}

\newcommand{\be}{\begin{equation}}
\newcommand{\ee}{\end{equation}}

\newcommand{\bea}{\begin{eqnarray}}
\newcommand{\eea}{\end{eqnarray}}

\newcommand{\bes}{\begin{subequations}}
\newcommand{\ees}{\end{subequations}}

\newcommand{\cN}{{\cal N}}

\def\sst#1{{\scriptscriptstyle #1}}

\def\0{{\sst{(0)}}}
\def\1{{\sst{(1)}}}
\def\2{{\sst{(2)}}}
\def\3{{\sst{(3)}}}
\def\4{{\sst{(4)}}}
\def\5{{\sst{(5)}}}
\def\6{{\sst{(6)}}}
\def\7{{\sst{(7)}}}
\def\8{{\sst{(8)}}}

\newcommand{\cW}{{\cal W}}

\newcommand{\vol}{\textrm{vol}}

\allowdisplaybreaks  %Introduces page breaks inside \eqnarray%

\usepackage{multirow}
\usepackage{rotating}

\def\z{\zeta_{12}}
\def\d{\textrm{d}}

\newcommand{\ba}{\begin{align}}
\newcommand{\ea}{\end{align}}

\newcommand{\bse}{\begin{subequations}}
\newcommand{\ese}{\end{subequations}}

\allowdisplaybreaks

\let\OLDthebibliography\thebibliography
\renewcommand\thebibliography[1]{
  \OLDthebibliography{#1}
  \setlength{\parskip}{4.46pt}
}

\begin{document}

\makeatletter
\renewcommand{\theequation}{\thesection.\arabic{equation}}
\@addtoreset{equation}{section}
\makeatother

\begin{titlepage}

\begin{flushright}
NIKHEF-2016-018 \\
\end{flushright}

\vspace{25pt}

   \begin{center}
   \baselineskip=16pt
   \begin{Large}\textbf{
Romans-mass-driven flows on the D2-brane}
   \end{Large}

\vspace{25pt}
		
{\large  Adolfo Guarino$^{1}$ ,\, Javier Tarr\'io$^{2}$ \,and\,  Oscar Varela$^{3,4}$}
		
\vspace{25pt}

	\begin{small}

	{\it $^{1}$ Nikhef Theory Group, Science Park 105, 1098 XG Amsterdam, The Netherlands.  } \\

	\vspace{15pt}
	
	{\it $^{2}$  Physique Th\'eorique et Math\'ematique, Universit\'e Libre de Bruxelles \\
	and International Solvay Institutes, ULB-Campus Plaine CP231, B-1050 Brussels, Belgium.} \\

	\vspace{15pt}
          
   {\it $^{3}$ Max-Planck-Institut f\"ur Gravitationsphysik (Albert-Einstein-Institut), \\Am M\"uhlenberg 1, D-14476 Potsdam, Germany.  } \\

	\vspace{15pt}
	
 {\it $^{4}$ Department of Physics, Utah State University, Logan, UT 84322, USA.} 

		\vspace{30pt}

 aguarino@nikhef.nl, jtarrio@ulb.ac.be, oscar.varela@aei.mpg.de

	\end{small}

\vskip 50pt

\end{center}

\begin{center}
\textbf{Abstract}
\end{center}

\begin{quote}

The addition of supersymmetric Chern-Simons terms to $\cN=8$ super-Yang-Mills theory in three-dimensions is expected to make the latter flow into infrared superconformal phases. We address this problem holographically by studying the effect of the Romans mass on the D2-brane near-horizon geometry. Working in a consistent, effective four-dimensional setting provided by $D=4$ $\cN=8$ supergravity with a dyonic ISO(7) gauging, we verify the existence of a rich web of supersymmetric domain walls triggered by the Romans mass that interpolate between the (four-dimensional description of the) D2-brane and various superconformal phases. We also construct domain walls for which both endpoints are superconformal. While most of our results are numerical, we provide analytic results for the $\textrm{SU}(3)\times \textrm{U}(1)$-invariant flow into an $\cN=2$ conformal phase recently discovered.

\end{quote}

\vfill

\end{titlepage}

\tableofcontents

%%%%%%%%%%%%%%%%%%%%%%%%%%%%%%%%%%%%%%%%%%%%%%%%%%%%%%%%%%%%%%%%%%%%%%%%%%

\section{Introduction}

Among the branes of string and M-theory, the D3, M2 and M5 branes enjoy a somewhat distinguished status in that, when considered in a flat background, their worldvolumes respectively support four- \cite{Maldacena:1997re}, three- \cite{Aharony:2008ug} and six-dimensional  maximally supersymmetric conformal field theories (CFTs). The first two cases are by now very well understood. It is also well known how to engineer D3 and M2 brane configurations and background geometries that support CFTs with less than maximal supersymmetry. In some cases, these conformal phases with reduced supersymmetry on the D3 and the M2 branes are known to be related to the corresponding maximally supersymmetric CFTs via renormalisation group (RG) flow.

For example, the $\cN=8$ CFT on a stack of $N$ planar M2 branes on flat space is given by a Chern-Simons theory with a product gauge group $\textrm{U}(N)_{k} \times \textrm{U}(N)_{-k}$ at (sufficiently low) Chern-Simons levels $k$ and $-k$, coupled to bifundamental matter with a quartic superpotential \cite{Aharony:2008ug}. Another conformal phase of the M2-brane field theory is known \cite{Benna:2008zy} that has only $\cN=2$ supersymmetry, the same gauge group, SU(3) flavour symmetry, U(1) R-symmetry and a sextic superpotential. The latter CFT turns out to arise  \cite{Benna:2008zy}  as the infrared (IR) fixed point of the RG flow caused by a perturbation of the $\cN=8$ phase \cite{Aharony:2008ug} with an $\cN=2$ SU(3)-invariant mass term for the bifundamentals. The near horizon region of both $\cN=8$ and $\cN=2$ M2-brane configurations develop AdS$_4 \times S^7$ geometries. In the former case, such configuration is simply the Freund-Rubin direct product solution \cite{Freund:1980xh} with the maximally supersymmetric and SO$(8)$-symmetric round metric on the seven-sphere. In the latter case, the product is warped and supported by a non-vanishing internal value of the $D=11$ four-form and a distorted metric on $S^7$ \cite{Corrado:2001nv}. This configuration preserves $\textrm{SU(3)} \times \textrm{U}(1)$ local symmetry, in agreement with the global symmetry of the dual CFT. Precision AdS$_4$/CFT$_3$  checks have been performed using this set-up. In particular, the free energy of the CFT, computed in \cite{Jafferis:2011zi} using localisation techniques \cite{Kapustin:2009kz,Jafferis:2010un,Hama:2010av}, perfectly matches the gravitational free energy of the AdS$_4$ solution.

The situation for the D$p$ branes of string theory with $ p \neq 3$ is fundamentally different. The $(p+1)$-dimensional worldvolume of $N$ coincident D$p$ branes on flat space supports maximally supersymmetric Yang-Mills (SYM) with gauge group SU$(N)$, but this theory is not conformal for any $p$ different from $3$. See \cite{Itzhaki:1998dd,Boonstra:1998mp,Kanitscheider:2008kd} for descriptions of the holographic dictionary in these maximally supersymmetric but non-conformal cases. From the gravity side, the lack of conformality is reflected by near horizon geometries of domain-wall, rather than AdS, type supported by a running dilaton. Remarkably enough, however, the D$p$ brane field theory can still flow in some cases into conformal phases with reduced or no supersymmetry. See for example \cite{Faedo:2015ula,Faedo:2015urf} for recent studies of this situation in various contexts. In this paper, we will fix $p=2$, corresponding to the D2-brane field theory. Specifically, we will study holographically how the ultraviolet (UV) description of the D2-brane worldvolume theory in terms of three-dimensional $\cN=8$ SYM is modified by the presence of a non-vanishing Romans mass.

The Romans mass $\hat F_\0$ \cite{Romans:1985tz} induces Chern-Simons couplings on the D2-brane worldvolume, and these are expected to trigger RG flows that drive the worldvolume theory into IR superconformal phases. In field theory terms, the addition of the Romans mass corresponds to augmenting three-dimensional $\cN=8$ SYM with Chern-Simons-matter terms, with the Chern-Simons coupling $k$ identified with the (quantised) Romans mass, $\hat F_\0 = k/(2\pi \ell_s)$, as in \cite{Gaiotto:2009mv}. This expectation was recently made more precise in \cite{Guarino:2015jca}. A specific deformation by $\hat F_\0$ was argued to make three-dimensional $\cN=8$ SYM flow into an $\cN=2$ superconformal phase described by a Chern-Simons-matter theory with a single gauge group SU$(N)$ at level $k$, flavour symmetry SU$(3)$, R-symmetry U(1) and cubic superpotential. This field theory is of the type first envisaged in \cite{Schwarz:2004yj} as potentially relevant for holography, and further studied in \cite{Gaiotto:2007qi,Minwalla:2011ma}. The (near horizon) gravity dual was identified \cite{Guarino:2015jca} to be of the form AdS$_4 \times S^6$, where the product is warped, the solution is supported by non-vanishing internal IIA forms and the metric on $S^6$ displays an $\textrm{SU}(3) \times \textrm{U}(1)$ isometry that matches the global symmetry of the CFT. The free-energy of this CFT on $S^3$ was computed by localisation and found to be in perfect agreement with the gravitational free energy of the dual geometry \cite{Guarino:2015jca}.

The effect of this particular $\cN=2$ deformation of the $\cN=8$ D2-brane field theory by the Romans mass is, thus, qualitatively similar to the mass deformation \cite{Benna:2008zy} of the $\cN=8$ M2-brane theory \cite{Aharony:2008ug}, with the crucial difference that only in the latter case is the UV field theory also conformal as the IR. Both superconformal IR phases in the M2 \cite{Benna:2008zy} and D2- \cite{Guarino:2015jca} brane field theories have the same global symmetries. Also, they have AdS$_4$ gravity duals in M-theory \cite{Corrado:2001nv} and massive type IIA string theory \cite{Guarino:2015jca} with qualitatively similar properties. In the M2-brane context, this and other related RG flows have been studied holographically \cite{Ahn:2000aq,Ahn:2000mf,Bobev:2009ms} using four-dimensional SO(8)-gauged supergravity \cite{deWit:1982ig}, and uplifted \cite{Corrado:2001nv,Ahn:2001kw} on $S^7$ to M-theory using the consistent truncation of \cite{deWit:1986iy}.

Recent developments now make the holographic study of RG flows of three-dimensional $\cN=8$ SYM triggered by the addition of Chern-Simons-matter terms accessible through similar gauged supergravity techniques. Massive type IIA supergravity \cite{Romans:1985tz} turns out to admit a consistent truncation on $S^6$ to maximal supergravity in four dimensions with $\textrm{ISO}(7) \equiv \textrm{CSO}(7,0,1) \equiv \textrm{SO}(7) \ltimes \mathbb{R}^7$ gauge group \cite{Guarino:2015jca,Guarino:2015vca}. The ISO(7) gauging is of the dyonic type discussed in \cite{Dall'Agata:2012bb,Dall'Agata:2014ita} (see also \cite{Inverso:2015viq}), with the magnetic gauge coupling $m$ identified with the Romans mass, $m = \hat F_0$ \cite{Guarino:2015jca}. The consistency of the $S^6$ truncation, together with the supergravity and holographic identities $m = \hat F_0$ and  $\hat F_\0 = k/(2\pi \ell_s)$, renders $D=4$ $\cN=8$ dyonically-gauged ISO(7) supergravity the natural framework to study the  effect  of Chern-Simons-matter terms on the large $N$ D2-brane field theory.

The complete $\cN=8$ dyonically-gauged ISO(7) supergravity theory was explicitly constructed in \cite{Guarino:2015qaa} using the embedding tensor formalism \cite{deWit:2002vt,deWit:2007mt}. Unlike the purely electric ISO(7) gauging \cite{Hull:1984yy}, its dyonic counterpart exhibits a rich structure of (AdS) vacua, both supersymmetric and non-supersymmetric. The supersymmetric vacua with at least residual SU(3) symmetry include $\cN=1$ vacua with G$_2$ \cite{Borghese:2012qm} and SU(3) \cite{Guarino:2015qaa} residual bosonic symmetries, and an $\cN=2$ vacuum with $\textrm{SU}(3) \times \textrm{U}(1) $ symmetry \cite{Guarino:2015jca}. In addition, the theory has an $\cN=3$ vacuum \cite{Gallerati:2014xra} with SO(4) symmetry. The  latter will not play a significant role in this work, as we will focus on supersymmetric RG flows that preserve at least SU(3) symmetry. By the consistency of the truncation, all these vacua uplift on the six-sphere to AdS$_4$ solutions of massive type IIA supergravity with the same symmetry and supersymmetry as the corresponding $D=4$ vacuum. The $\cN=1$, G$_2$ massive IIA solution was first constructed, directly in $D=10$ by other methods, in \cite{Behrndt:2004km}. All other solutions were recently obtained by direct uplift using the formulae of \cite{Guarino:2015jca,Guarino:2015vca}. The $\cN=2$, $\textrm{SU}(3) \times \textrm{U}(1) $ vacuum uplifts \cite{Guarino:2015jca} to the $\cN=2$ massive IIA solution discussed above, and the $\cN=1$, SU(3) and $\cN=3$, SO(4) vacua were respectively uplifted in \cite{Varela:2015uca} and \cite{Pang:2015vna} (see also \cite{Pang:2015rwd}).

\begin{figure}[t!]
\begin{center}
\includegraphics[width=110mm,angle=0]{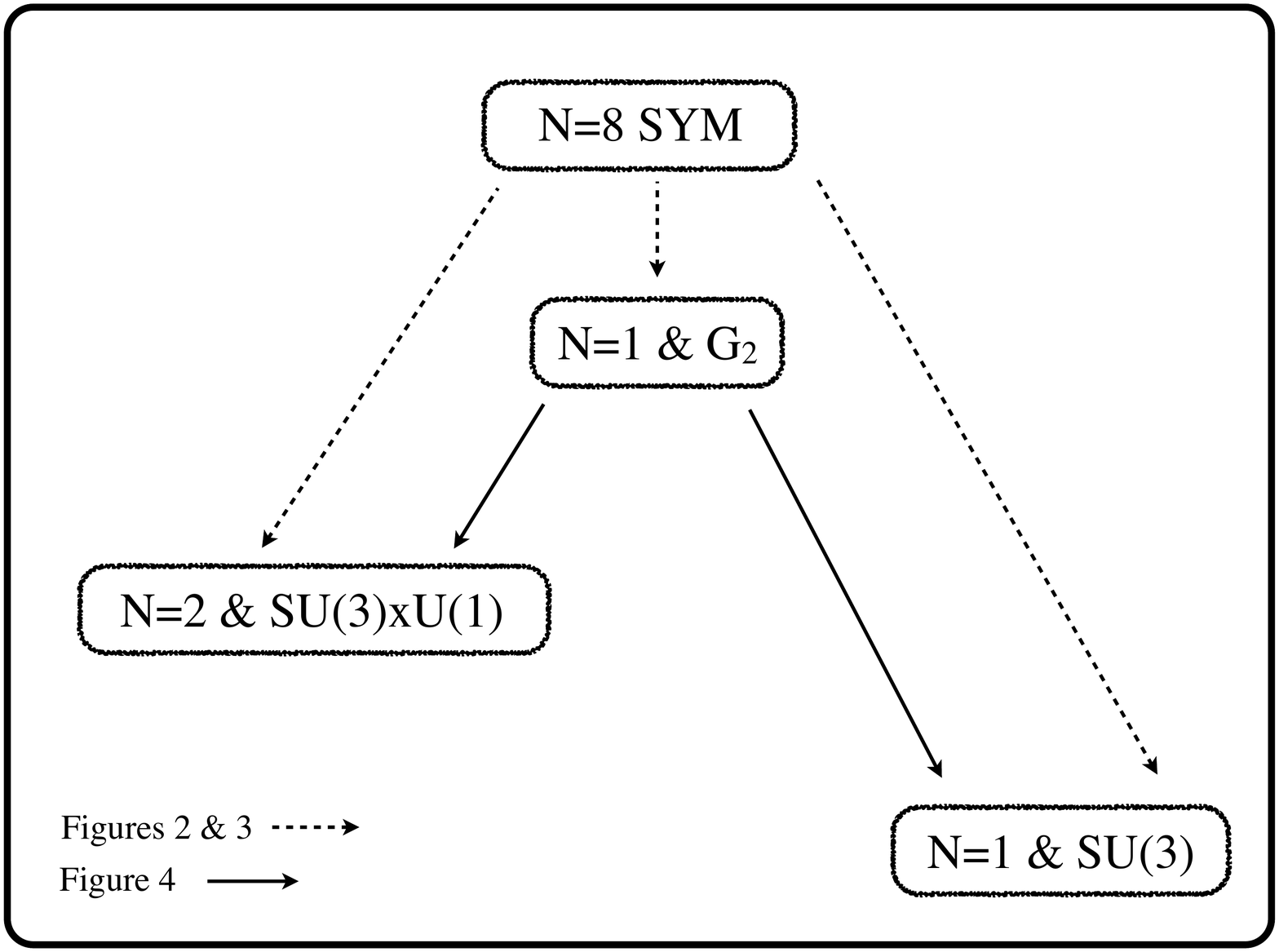}
\caption{Schematic representation of the holographic supersymmetric RG flows covered in this paper: from $\mathcal{N}=8$ SYM to IR CFTs (dotted lines) and between CFTs (solid lines). See figures \ref{fig.SYMtoG2domainwall},  \ref{fig.cones} and \ref{fig.AdSdomwalls} for actual plots.}
\label{Fig:DWs} 
\end{center}
\end{figure}

All these supersymmetric AdS$_4$ solutions of massive IIA string theory should correspond to conformal phases of the D2-brane field theory with distinct flavour symmetries and supersymmetry. They should arise as the IR endpoints of RG flows triggered by different symmetry- and supersymmetry-preserving deformations of $\cN=8$ SYM  by Chern-Simons-matter terms. We confirm this expectation for the $\cN=2$ flow discussed in \cite{Guarino:2015jca} by explicitly constructing a domain wall solution of $D=4$ dyonic ISO(7) supergravity that  interpolates between the ($D=4$ description of the) planar D2-brane solution in the UV and the $\cN=2$, $\textrm{SU}(3) \times \textrm{U}(1)$ vacuum in the IR. More generally, we show that there exists an entire family of supersymmetric SU(3)-invariant flows that originate in $\cN=8$ SYM and drive the theory towards the $\cN=2$, $\textrm{SU}(3) \times \textrm{U}(1)$-symmetric IR fixed point. We find a second family of supersymmetric RG flows that drive $\cN=8$ SYM into the $\cN=1$ IR phase with SU(3) invariance. Both families are bounded by a unique flow with IR endpoint in the $\cN=1$ G$_2$-symmetric phase. We also find two unique domain walls that interpolate between this G$_2$ conformal phase in the UV and either the $\cN=2$, $\textrm{SU}(3) \times \textrm{U}(1)$ point or the $\cN=1$ SU(3) point in the IR. By the generic results of \cite{Guarino:2015jca,Guarino:2015vca} and the specific formulae of \cite{Varela:2015uca}, all these domain walls uplift to massive type IIA supergravity and link the corresponding AdS$_4$ solutions. See figure~\ref{Fig:DWs} for a sketch of this web of domain walls.   

In section \ref{sec:SU3sector} we review the natural arena for the holographic RG flows that we construct in this paper: the SU(3)-invariant sector of the $\cN=8$ dyonic supergravity. We discuss the AdS vacuum structure and the flow equations. The flows that interpolate between the D2-brane behaviour and the IR conformal phases with at least SU(3) flavour symmetry are constructed in section \ref{sec:SYMtoCFT}. Section \ref{CFTtoCFT} deals with flows between conformal phases, and section \ref{sec:Discussion} provides further discussion. Various appendices close the paper. Appendix \ref{App:boundary_conds} elaborates on the boundary conditions satisfied by our domain walls, appendix \ref{App:RunningFE} comments on the running of the free energy along the flows and appendix \ref{App:SO4} constructs some flows into the $\cN=3$ SO(4)-invariant conformal phase.

%%%%%%%%%%%%%%%
%%%%%%%%%%%%%%%

\section{The SU(3)-invariant sector of dyonic ISO(7) supergravity } \label{sec:SU3sector}

We want to construct supersymmetric domain walls of $D=4$ $\cN=8$ dyonic ISO(7)-gauged supergravity that preserve at least $\textrm{SU}(3)\subset \textrm{ISO}(7)$ symmetry. The natural venue to look for such solutions is, thus, the SU(3)-invariant sector of the $D=4$ supergravity. This was explicitly worked out in \cite{Guarino:2015qaa}. The SU(3)-invariant sector is $\cN=2$ supersymmetric and contains one vector multiplet and one hypermultiplet and, accordingly, six real scalars along with vectors and other tensor fields in the SU(3)-invariant tensor hierarchy. Here we will be only interested in neutral domain wall solutions. For this reason, we consistently truncate out the vectors and work only with the four neutral scalars of the theory, together with the metric.

\subsection{Flow equations and fixed points} \label{subsec:Floweqs}

We find it useful to use the parameterisation introduced in section 3.3 of \cite{Guarino:2015qaa} following \cite{Bobev:2009ms}. The four neutral real scalars are thereby packed into two complex scalars $z$ and $\zeta_{12}$ which take values on two copies of the Poincar\'e unit disk: $\,|z|<1\,$ and $\,|\z|<1\,$. These respectively correspond to the scalars in the vector multiplet and the neutral scalars in the hypermultiplet. The Einstein-scalar action reads \cite{Guarino:2015qaa}
\ba
S & = \frac{1}{16\pi G_4} \int \d^4 x\, \sqrt{-g} \Bigg[ R - \frac{6}{(1-|z|^2)^2}\partial_\mu z \, \partial^\mu \bar z - \frac{8}{(1-|\z|^2)^2}\partial_\mu \z \, \partial^\mu \bar{\zeta}_{12} -V  \Bigg] \ , \label{eq.SU3action}
\end{align}
where the scalar potential $\,V\,$ can be written as 
\ba
V & =  2 \left( \frac{4}{3} (1-|z|^2)^2 \left| \frac{\partial W}{\partial z} \right|^2  +  (1-|\z|^2)^2 \left| \frac{\partial W}{\partial \z} \right|^2  - 3 W^2 \right) \ , \label{eq.SU3Potential}
\end{align}
in terms of either of two (real) superpotentials,  $\,W=|\cW(z,\zeta_{12})|\,$ or $\,W=|\cW(z,\bar{\zeta}_{12})|\,$, with \cite{Guarino:2015qaa} 
\ba
\label{eq.WSU3}
\cW(z,\zeta_{12}) & = 2\,   (1-|z|^{2})^{-\frac{3}{2}} \, (1-|\zeta_{12}|^{2})^{-2} \nonumber \\
& \qquad \times   \Bigg[ \, g \, \left( \dfrac{7}{8} \,  (1- \zeta_{12})^{4} \, (1+z)^{3} \, + \, 3 \, (\zeta_{12}-z) \, (1+z) \,(1-\zeta_{12})^2 \,(1- z \, \zeta_{12})  \right)  \nonumber \\
& \qquad \qquad +   i\, \dfrac{m}{8} \,  (1-\zeta_{12})^{4} \, (1-z)^{3}  \, \Bigg] \ .
\end{align}
Here, $\,g\,$ and $\,m \equiv g c\,$ respectively are the electric and magnetic couplings of ISO(7) supergravity, and $c$ the `dyonically gauging parameter'. As explained in \cite{Dall'Agata:2014ita}, all theories with $c\neq0$ are classically equivalent. Accordingly, $\,c\,$ can be fixed to any (non-zero) value without loss of generality. Note however that the position of the fixed points in scalar-space, and therefore the domain walls connecting them, are \mbox{$c$-dependent} in this parameterisation. Upon truncation from massive type IIA, $g$ becomes proportional to the inverse $S^6$ radius and $m$ identified with the Romans mass \cite{Guarino:2015jca}.

Similar superpotentials in the SU(3)-invariant sector of related $\cN=8$ gaugings have been previously constructed in \cite{Ahn:2000mf,Bobev:2009ms,Borghese:2012zs,Guarino:2015tja}. For superpotentials in the SO$(7)_+$ and G$_2$ invariant sectors of the purely electric ($c=0$, $m=0$) $\cN=8$ ISO(7) gauging \cite{Hull:1984yy}, see \cite{Ahn:2001by,Ahn:2002qga}. The latter reference allows us to crosscheck the purely electric, G$_2$-invariant truncation of our superpotential: setting $m=0$ and $\zeta_{12} = z$ in (\ref{eq.WSU3}) we reproduce the superpotential $z_2$ given in equation ($14$) of \cite{Ahn:2002qga} after the field redefinition $\,z=\tanh \big(\lambda/(2\sqrt{2} ) \big) \, e^{-i\alpha}\,$. More generally, we have verified that $\cW(z,\zeta_{12})$ and $\cW(z,\bar{\zeta}_{12})$ arise as the two SU(3)-invariant eigenvalues at a generic point in scalar space of the full $\cN=8$ gravitino mass matrix of dyonic ISO(7) supergravity.

We are interested in RG flows that preserve some supersymmetry on the D2-brane. Holographically, these correspond to domain wall solutions to the equations of motion that follow from (\ref{eq.SU3action}) for which the metric takes on the local form
\be
\label{eq.dwmetric}
\d s_{4}^2 = e^{2A(r)} \, \eta_{\alpha\beta} \, \d x^{\alpha} \, \d x^{\beta} + \d r^2 \; , 
\hspace{6mm} \textrm{ where } \hspace{6mm}
\eta_{\alpha\beta} = \textrm{diag}(-1,+1,+1) \ .
\ee
The scale factor $A(r)$ and the complex scalars $\,z(r)\,$, $\,\zeta_{12}(r)\,$  depend only on the coordinate $r$ transverse to the three flat directions $x^\alpha$, $\alpha = 0, 1,2$. The domain walls will be supersymmetric provided the supersymmetry variations of the fermions vanish. Selecting henceforth $\,W=|\cW(z,\zeta_{12})|\,$ for definiteness,  this turns out to be equivalent to the following set of first order BPS equations:
\be
\label{eq.BPSeqs}
\frac{dz}{dr} = -\frac{2}{3}\,  (1-|z|^2)^2 \, \frac{\partial W}{\partial \bar{z}} 
\hspace{5mm}  ,   \hspace{5mm} 
\frac{d\z}{dr} = - \frac{1}{2} \, (1-|\z|^2)^2 \,  \frac{\partial W}{\partial \bar{\zeta}_{12}}
\hspace{5mm}  ,   \hspace{5mm} 
\frac{dA}{dr} = W \ .
\ee
A generic feature of these equations is that, if a solution to the first two is found for the scalars $z(r)$ and $\zeta_{12}(r)$, then the scale factor equation can be integrated upon substitution of those scalar profiles into the superpotential $W$. 

The derivation of (\ref{eq.BPSeqs}) from the supersymmetry variations parallels \cite{Ahn:2000aq,Ahn:2000mf,Ahn:2002qga}. Turning on the dyonic parameter $\,c\,$ in an SU(3)-invariant manner leads to new, $m$-dependent terms in the fermion shift matrices $\,A_1\,$, $\,A_2\,$ of the $\cN=8$ supergravity, but does not turn on additional components with respect to the electric ISO(7) gauging. Also, important purely electric expressions, like ($25$) of \cite{Ahn:2000mf} and ($24$) of \cite{Ahn:2002qga} still hold at $\,c \neq 0\,$. These allow us to write the BPS equations (\ref{eq.BPSeqs}) in terms of the real superpotential $W$, rather than the complex (\ref{eq.WSU3}). As in the cases previously dealt with in the literature, projections on the Killing spinor need to be imposed in order to obtain (\ref{eq.BPSeqs}) from the supersymmetry variations. Accordingly, generic domain wall solutions to these equations and their dual field theory flows will generically preserve two real (Poincar\'e) supercharges. We will also find domain walls, and their dual flows, that preserve four real supercharges. 

\label{Ncomments} Some remarks about terminology are now in order. We usually adhere to the standard, though perhaps ambiguous, convention of denoting by $\cN$ the total number of fermionic generators in the bulk, but only the superPoincar\'e generators in the boundary excluding any superconformal generators, if present. Thus, we simultaneously speak of $D=4$ $\cN=8$ supergravity (32 supercharges) and three-dimensional $\cN=8$ SYM (16 supercharges). However, also following standard practice, we will use the field theory convention ({\it i.e.} we will use `the $\cN$ of the boundary') to refer to the two- and four-supercharge domain walls and their dual flows as $\cN=1$ and $\cN=2$. Only at a fixed point of the flow equations, to be discussed shortly, $\cN$ of both bulk and boundary coincide numerically due to the presence of additional superconformal charges in the latter. For example, the $\cN=2$ AdS fixed point (eight supercharges) is dual to the $\cN=2$ superconformal field theory (four Poincar\'e and four superconformal supercharges) discussed in \cite{Guarino:2015jca}.

The BPS equations (\ref{eq.BPSeqs}) have three fixed points, \emph{i.e.}, solutions with constant ($r$-independent) scalars $(z_*,{\z}_*)$, corresponding to extrema $\,W_*\equiv W(z_*,{\z}_*)\,$ of the superpotential \cite{Guarino:2015qaa}. These vacua are AdS, as can be easily seen by solving the last equation in (\ref{eq.BPSeqs}),
\be \label{ScaleFactorAdS}
A(r) = W_*\,r \equiv \frac{r}{L_*} \; ,
\ee
whereby (\ref{eq.dwmetric}) becomes the usual AdS metric in the Poincar\'e patch upon the coordinate redefinition $\,\rho=L_* \,e^{r/L_*}\,$. In (\ref{ScaleFactorAdS}), $\,L^2_* \equiv -6/V_{*}\,$ is the squared AdS radius, with $\,V_*\equiv V(z_*,{\z}_*) < 0\,$ the scalar potential at the extremum, namely, the corresponding cosmological constant. The boundary  corresponds to $r\to\infty$ and the Poincar\'e horizon to $r\to-\infty$.

\begin{table}[t]
\renewcommand{\arraystretch}{1.5}
\scalebox{0.86}{
\begin{tabular}{@{}c|cccc@{}}
\toprule
& $z_*$ & ${\z}_*$ & $L_*$ & $V_*$ \\ \midrule
\multirow{2}{*}{\makecell{$\cN=2$ \\[1mm] SU(3)$\times$U(1)}}    & \multirow{2}{*}{$\dfrac{-2+(\sqrt{3}-i)c^{1/3}}{2+(\sqrt{3}-i)c^{1/3}} $} & \multirow{2}{*}{$\dfrac{-2+\sqrt{2}\, c^{1/3}}{2+\sqrt{2}\, c^{1/3}}$} & \multirow{2}{*}{$\dfrac{1}{\sqrt{2}\,3^{1/4}} \dfrac{c^{1/6}}{g}$} & \multirow{2}{*}{$-12\sqrt{3} \, \dfrac{g^2}{c^{1/3}}$}   \\[1mm]
 &   &  &   &  \\ \midrule
\multirow{2}{*}{\makecell{$\cN=1$ \\[1mm] G$_2$}}     & \multirow{2}{*}{$\dfrac{-8+2^{2/3}(\sqrt{15}-i)c^{1/3}}{8+2^{2/3}(\sqrt{15}-i)c^{1/3}}$} & \multirow{2}{*}{$\dfrac{-8+2^{2/3}(\sqrt{15}-i)c^{1/3}}{8+2^{2/3}(\sqrt{15}-i)c^{1/3}}$} & \multirow{2}{*}{$\dfrac{5 \cdot 15^{1/4}}{16\cdot 2^{1/6}} \dfrac{c^{1/6}}{g}$} & \multirow{2}{*}{$-\dfrac{512\cdot 2^{1/3}}{25}\sqrt{\dfrac{3}{5}} \dfrac{g^2}{c^{1/3}}$}   \\[1mm]
 &   &  &   &  \\ \midrule  %
\multirow{2}{*}{\makecell{$\cN=1$ \\[1mm] SU(3)}}    & \multirow{2}{*}{$\dfrac{-4+(\sqrt{15}+i)c^{1/3}}{4+(\sqrt{15}+i)c^{1/3}}$} & \multirow{2}{*}{$\dfrac{-4+(\sqrt{5}-i\sqrt{3})c^{1/3}}{4+(\sqrt{5}-i\sqrt{3})c^{1/3}}$} & \multirow{2}{*}{$\dfrac{5^{5/4}}{8\sqrt{2}\, 3^{1/4}} \dfrac{c^{1/6}}{g}$} & \multirow{2}{*}{$-\dfrac{768}{25}\sqrt{\frac{3}{5}} \dfrac{g^2}{c^{1/3}}$}   \\[1mm]
 &   &  &   &  \\   \bottomrule
\end{tabular}
}
\caption{Supersymmetric fixed points within the SU(3)-invariant sector of $D=4$ $\,\cN=8\,$ dyonic ISO(7)-gauged supergravity. These correspond to AdS solutions with radius $L_*$ and cosmological constant $V_*$, and depend on the dimensionless ratio $c\equiv m/g$.}
\label{tab.SU3fps}  
\end{table}

The total symmetry of each AdS fixed point is $\textrm{OSp}(4|\cN) \times G$, where $\cN=1$ or $\cN=2$ labels the residual supersymmetry, and $G= \textrm{G}_2$ or $G= \textrm{SU}(3)$ for $\cN=1$ and $G= \textrm{SU}(3) \times \textrm{U}(1)$ for $\cN=2$ denotes the residual bosonic symmetry. This is the local symmetry that these fixed points preserve both as solutions \cite{Borghese:2012qm,Guarino:2015qaa,Guarino:2015jca} of $D=4$ $\cN=8$ dyonic ISO(7) supergravity and as solutions \cite{Behrndt:2004km,Varela:2015uca,Guarino:2015jca} of massive type IIA. This is also the global symmetry of the dual CFTs. In particular, $G$ corresponds to the flavour symmetry in the $\cN=1$ cases. In the $\cN=2$ case, the SU(3) and U(1) factors respectively correspond to the flavour and the R-symmetry. As already noted in \cite{Guarino:2015qaa}, in the purely electric \cite{Hull:1984yy}, $c \rightarrow 0$, or purely magnetic, $c \rightarrow \infty$, limits the extrema disappear from the proper Poincar\'e disks: $|z_*| \rightarrow 1$, $|\zeta_{12*}| \rightarrow 1$. Thus, these AdS solutions only exist  for the dyonic ISO(7) gauging. See table 1 for the location of the fixed points in the $\,(z,\z)\,$ parameterisation.

%%%%%%%%%%%%%%%%%
%%%%%%%%%%%%%%%%%

\subsection{Modes and dimensions of dual operators}

%%%%%%%%%%%%%%%%%
%%%%%%%%%%%%%%%%%

We are interested in constructing supersymmetric domain wall solutions dual to RG flows with at least one endpoint at one of the fixed points collected in table \ref{tab.SU3fps}. Let us show that the set of SU(3)-invariant perturbations about these points include the modes necessary to describe the relevant and irrelevant directions of the dual RG flows. This analysis allows us to determine the relevant boundary conditions for the integration of  (\ref{eq.BPSeqs}).

As we have already noted, scalars and metric scale factor decouple in the flow equations. Accordingly,  for this analysis we can simply focus on solving for the scalars only. The linearisation of \eqref{eq.BPSeqs}  around any AdS fixed point has a general solution given by the linear superposition
\be
\label{eq.perturbAdS}
z-z_* = \sum_{i=1}^4 a_i \, e^{-\tilde\Delta_i \frac{r}{L_*}} 
\hspace{8mm} , \hspace{8mm}
\z-{\z}_* = \sum_{i=1}^4 \alpha_i \, e^{-\tilde\Delta_i \frac{r}{L_*}} \ .
\ee
Here, $a_i$ and $\alpha_i$ are eight complex constants that can be written in terms of only four independent real parameters  in one-to-one correspondence with the exponents $\tilde \Delta_i$, see appendix \ref{App:boundary_conds}. The latter turn out to coincide with one of the two roots, $\,\Delta_{\pm}\,$, of the equation
\be
\label{eq.deltaeq}
M^2L_*^2=\Delta(\Delta-3) \ ,
\ee
that holographically relates the mass of a bulk scalar field to the conformal dimension of its dual operator in the  boundary. The dual operators have conformal dimensions given by the largest root $\Delta_{+}$ and correspond to relevant (irrelevant) deformations of the CFT if $\Delta_{+}<3\,$ ($\Delta_{+}>3$). When the masses of the scalar perturbations  around the AdS fixed point  lie in the range ${\,-9/4<M^2L_*^2<-5/4\,}$ an alternative quantisation is possible, and the mode can describe a dual operator with dimension $\Delta_-$. In table~\ref{tab.deltas} we import from  \cite{Guarino:2015qaa} the SU(3)-invariant scalar mass spectrum around each of the fixed points, along with the two solutions $\,\Delta_{\pm}\,$ to \eqref{eq.deltaeq}. The specific $\tilde \Delta_i$ that appear in the linearised solution (\ref{eq.perturbAdS}) are highlighted with a gray background.

\begin{table}[t]
\center
\renewcommand{\arraystretch}{1.5}
\scalebox{0.89}{
\begin{tabular}{@{}cr|cccc|cc@{}}
\toprule
&& Mode 1 & Mode 2 & Mode 3 & Mode 4 & Relevant oper. & Irrelevant oper.  \\ \midrule
\multirow{3}{*}{\makecell{$\cN=2$ \\[1mm] SU(3)$\times$U(1)}} &
   $M^2L_*^2$    & $3-\sqrt{17}$ & $2$ & $2$ & $3+\sqrt{17}$   \\[1mm]
&  $\Delta_+$    & \cellcolor{gray!10} ${\color{blue}\mathbf{\frac{1+\sqrt{17}}{2}}}$  & \cellcolor{gray!10}  ${\color{blue}\mathbf{\frac{3+\sqrt{17}}{2}}}$  & $\frac{3+\sqrt{17}}{2}$  & $\frac{5+\sqrt{17}}{2}$ & 1 & 3 \\
&  $\Delta_-$    & $\frac{5-\sqrt{17}}{2}$ & $\frac{3-\sqrt{17}}{2}$  & \cellcolor{gray!10}  ${\color{red}\mathbf{\frac{3-\sqrt{17}}{2}}}$  & \cellcolor{gray!10}  ${\color{red}\mathbf{\frac{1-\sqrt{17}}{2}}}$  \\ \midrule
\multirow{3}{*}{\makecell{$\cN=1$ \\[1mm] G$_2$}} &
   $M^2L_*^2$    & $\frac{-11-\sqrt{6}}{6}$ & $\frac{-11+\sqrt{6}}{6}$ & $4-\sqrt{6}$ & $4+\sqrt{6}$   \\[1mm]
&  $\Delta_+$    & $2-\frac{1}{\sqrt{6}}$ & $2+\frac{1}{\sqrt{6}}$  & \cellcolor{gray!10}  ${\color{blue}\mathbf{1+\sqrt{6}}}$  & $2+\sqrt{6}$  & 2 & 2 \\
&  $\Delta_-$    & \cellcolor{gray!10}  ${\color{blue}\mathbf{1+\frac{1}{\sqrt{6}}}}$ &  \cellcolor{gray!10} ${\color{blue}\mathbf{1-\frac{1}{\sqrt{6}}}}$  & $2-\sqrt{6}$   & \cellcolor{gray!10}  ${\color{red}\mathbf{1-\sqrt{6}}}$    \\ \midrule
\multirow{3}{*}{\makecell{$\cN=1$ \\[1mm] SU(3)}} &
   $M^2L_*^2$    & $4-\sqrt{6}$ & $4-\sqrt{6}$ & $4+\sqrt{6}$  & $4+\sqrt{6}$   \\[1mm]
&  $\Delta_+$    & \cellcolor{gray!10} ${\color{blue}\mathbf{1+\sqrt{6}}}$  & \cellcolor{gray!10}  ${\color{blue}\mathbf{1+\sqrt{6}}}$  & $2+\sqrt{6}$  & $2+\sqrt{6}$  & 0 & 4 \\
&  $\Delta_-$    & $2-\sqrt{6}$   & $2-\sqrt{6}$   & \cellcolor{gray!10}   ${\color{red}\mathbf{1-\sqrt{6}}}$   & \cellcolor{gray!10}  ${\color{red}\mathbf{1-\sqrt{6}}}$    \\
  \bottomrule
\end{tabular}
}
\caption{Scalar mass spectrum around each of the AdS fixed points of table \ref{tab.SU3fps}. The roots $\tilde \Delta$ (either $\tilde \Delta = \Delta_+$ or $\tilde \Delta = \Delta_-$) of equation (\ref{eq.deltaeq})  that appear in the linearised solution \eqref{eq.perturbAdS} are shown with a gray background. Blue (red) values are compatible with regularity of a domain wall that reaches the corresponding fixed point in the UV (IR).  The last two columns indicate the number of relevant and irrelevant operators at each point.}
\label{tab.deltas}  
\end{table}

From \eqref{eq.perturbAdS} it can be seen that a regular domain wall must approach a UV ($ r \rightarrow + \infty$) or IR ($ r \rightarrow - \infty$) fixed point driven by modes with $\tilde{\Delta} >0$ or $\tilde{\Delta}  < 0$, respectively. Table \ref{tab.deltas} graphically shows these signs with a self-explanatory colour code: blue in the first case and red in the second. It is apparent from the table that all these fixed points can serve as either UV or IR endpoints of domain walls. However, only the G$_2$ fixed point will happen to realise this feature in this paper. For all our flows it also happens, consistently enough, that the active modes correspond to relevant or irrelevant operators in the field theory depending on whether the fixed point serves as a UV or IR phase. Finally, only non-normalisable fall-offs  will turn out to be activated in our flows, {\it i.e.},  $\tilde{\Delta} = \Delta_- $ always. This confirms the expectation that our domain walls are dual to RG flows caused by perturbations of the field theory, rather than vacuum expectation values.

In the next two sections we numerically integrate the BPS equations (\ref{eq.BPSeqs}) with the boundary conditions specified in table \ref{tab.deltas} and appendix \ref{App:boundary_conds}. We find two types of regular domain walls. The first type corresponds to solutions for which one of the superconformal fixed points lies at the IR end, while the UV is dominated by the non-conformal $\cN=8$ SYM theory. These are flows of $\cN=8$ SYM that are generated upon perturbation with supersymmetric  Chern-Simons-matter terms. We subsequently refer to these solutions as `SYM to CFT flows'. The second kind corresponds to domain walls connecting two fixed points, and we refer to them as `CFT to CFT flows'. Similar supersymmetric flows of the latter type in the purely electric SO(8) gauging \cite{deWit:1982ig} of $D=4$ $\cN=8$ supergravity have been previously constructed in \cite{Ahn:2000aq,Ahn:2000mf,Corrado:2001nv,Bobev:2009ms} and, in the dyonic SO(8) gauging \cite{Dall'Agata:2012bb}, in \cite{Guarino:2013gsa,Tarrio:2013qga,Pang:2015mra}.

%%%%%%%%%%%%%%%%%
%%%%%%%%%%%%%%%%%

\section{SYM to CFT flows} \label{sec:SYMtoCFT}

Let us first discuss the holographic RG flows that originate upon modifying $\cN=8$ SYM with Chern-Simons-matter  terms. As we discussed in the introduction, this holographically corresponds to perturbing the D2-brane worldvolume theory with different couplings governed by the Romans mass $\hat F_\0$.

\subsection{Generalities}

Our starting point is a stack of $N$ D2-branes of massless type IIA string theory in flat space. In the type IIA conventions of appendix A of \cite{Guarino:2015vca}, the near horizon region of such configuration reads, in  Einstein frame,
\begin{equation}
\label{D2embedding}
\begin{array}{rll}
d \hat{s}_{10}^2 &=& e^{\frac{3}{4} \varphi(r)} \,  \big(e^{2A(r)} \, \eta_{\alpha\beta} \, \d x^{\alpha} \, \d x^{\beta} + \d r^2 \big) + g^{-2} e^{-\frac{1}{4} \varphi(r)} \,  ds^2(S^6) \ ,  \\[5pt]
e^{\hat \phi} &=& e^{\frac{5}{2} \varphi(r)} \ , \\[5pt]
\hat F_\4 &=& \frac{5}{3!} \, g \, e^{\varphi(r)+3A(r)} \,  \epsilon_{\alpha \beta \gamma} \, \d x^\alpha \wedge \d x^\beta \wedge \d x^\gamma \wedge \d r   \ , \\[5pt]
\hat H_\3 &=& \hat F_\2 = 0 \ .
\end{array}
\end{equation} 
Here $ds^2(S^6)$ is the conventional, round, SO(7)-symmetric Einstein metric on the six-sphere, normalised so that the Ricci tensor equals five times the metric, and 
\be \label{phiA}
e^{\varphi(r)}= \frac{2^4}{  (g \, r)^2} \quad  , \qquad e^{A(r)} = (g\,r)^7 \; . 
\ee
We have found it useful to write this near horizon D2-brane solution in terms of a constant $g$. This is related to the inverse radius $L$ of $S^6$ as $g=1/L$. The latter is in turn related upon flux quantisation to the number $N$ of D2-branes\footnote{See \cite{Itzhaki:1998dd,Kanitscheider:2008kd} for further details. Moving from the Einstein to the string frame and changing coordinates as $g r=4 (u/L)^{1/4}$ brings the massless IIA solution (\ref{D2embedding}) to the form presented in \cite{Itzhaki:1998dd}. More concretely, we have $\,e^{\hat{\phi}_{\textrm{there}}}=g_{YM}^{2} \, e^{\hat{\phi}_{\textrm{here}}}\,$ with $\,g_{YM}^{2}N=L^5/(6\pi^2)\,$ and $\,x^{\alpha}_{\textrm{there}}=2^{-14} x^{\alpha}_{\textrm{here}}\,$.}. The near-horizon solution (\ref{D2embedding}) is $1/2$-BPS, {\it i.e.}, preserves sixteen supercharges, and is manifestly invariant under SO(7), the R-symmetry group of three-dimensional $\cN=8$ SYM. It takes on a warped product form of the round metric on $S^6$ and a four-dimensional domain wall metric of the type (\ref{eq.dwmetric}). 

A natural counterpart of the solution (\ref{D2embedding}) in M-theory is provided by the SO(8)-invariant direct product Freund-Rubin solution $\textrm{AdS}_4 \times S^7$ \cite{Freund:1980xh} that arises as the near-horizon limit of a stack of M2-branes. In that $D=11$ case, the external domain wall metric is promoted to the usual metric on the Poincar\'e patch of AdS$_4$ and the number of supersymmetries is accordingly enhanced to include sixteen additional superconformal ones. The consistent truncation of $D=11$ supergravity on the seven-sphere \cite{deWit:1986iy} down to $D=4$ $\cN=8$ (electrically-gauged) SO(8) supergravity \cite{deWit:1982ig} can be used to factor out the $S^7$ dependence and work consistently in an effective four-dimensional setting. From this point of view, the Freund-Rubin solution corresponds to the `central' $\cN=8$, SO(8)-invariant AdS stationary point of the $D=4$ gauged supergravity. The scalars of the $D=4$ theory can be interpreted as couplings in the dual large-$N$ M2-brane field theory of \cite{Aharony:2008ug}. When turned on, these can trigger RG flows into other IR conformal phases with less symmetry and supersymmetry. From the effective four-dimensional perspective, these IR phases correspond to other AdS critical points of the scalar potential of the SO(8)-gauged supergravity, and the RG flows are manifestly exhibited holographically as domain walls between the AdS fixed points \cite{Ahn:2000aq,Ahn:2000mf,Corrado:2001nv,Bobev:2009ms}. By the consistency of the $S^7$ truncation \cite{deWit:1986iy}, there exists a one-to-one correspondence between four-dimensional and eleven-dimensional solutions, be them AdS vacua or domain walls.

An analogue picture emerges in our present D2-brane context, with some similarities and various crucial differences. Similarly to the $D=11$ case, both massless and massive type IIA supergravity can be consistently truncated on the six-sphere to $D=4$ $\cN=8$ supergravity with an $\textrm{ISO}(7) \equiv \textrm{CSO}(7,0,1) \equiv \textrm{SO}(7) \ltimes \mathbb{R}^7$  gauging. In the massless case, the gauging is purely electric and was constructed long ago \cite{Hull:1984yy}. The consistency of the truncation was first suggested in \cite{Hull:1988jw} and recently made more precise in \cite{Guarino:2015jca,Guarino:2015vca}. In the massive case, the ISO(7) gauging is dyonic, in the sense of  \cite{Dall'Agata:2012bb,Dall'Agata:2014ita,Inverso:2015viq}. The dyonic four-dimensional supergravity was constructed in \cite{Guarino:2015qaa}, and the consistency of the $S^6$ truncation shown in\footnote{Maximally supersymmetric truncations of massive type IIA \cite{Romans:1985tz} on $S^n$ appear to be inconsistent for all $n \leq 6$ different from the usual Scherk-Schwarz $n=1$ case and the $n=6$ case of \cite{Guarino:2015jca,Guarino:2015vca}, see \cite{Ciceri:2016dmd,Cassani:2016ncu}.} \cite{Guarino:2015jca,Guarino:2015vca}. 

Unlike the SO(8) gauging \cite{deWit:1982ig}, the purely electric ISO(7) gauging \cite{Hull:1984yy} has no (AdS) critical points. Stationary points with at least residual SO(7) and G$_2$ symmetry were respectively ruled out in  \cite{Hull:1984yy} itself and  \cite{Ahn:2002qga}. More recently, critical points of the electric ISO(7) gauging were excluded in general  in \cite{DallAgata:2011aa,Dall'Agata:2014ita,Guarino:2015qaa}. Thus, while the Freund-Rubin solution \cite{Freund:1980xh} corresponding to the near-horizon geometry of the M2-brane descends, upon truncation on $S^7$, to the $\cN=8$ SO(8) point of the electric SO(8) gauging \cite{deWit:1982ig} of $D=4$ $\cN=8$ supergravity, the same thing does not happen for the near horizon D2-brane solution (\ref{D2embedding}), (\ref{phiA}). Instead, as anticipated in \cite{Boonstra:1998mp}, the latter reduces on $S^6$ to a domain wall solution of $D=4$ $\cN=8$ electrically-gauged ISO(7) supergravity. This domain wall preserves sixteen out of the thirty-two supercharges of the $\cN=8$ supergravity, and the SO(7) subgroup of ISO(7). Since $\textrm{SU}(3) \subset \textrm{SO}(7)$, this solution is also contained in the SU(3)-invariant sector of the purely electric ISO(7) gauging, whose Lagrangian and flow equations are respectively given by (\ref{eq.SU3action})--(\ref{eq.WSU3}) and (\ref{eq.BPSeqs}) with $m \equiv gc = 0$, $g \neq 0$. In our conventions, this domain wall is given by the metric (\ref{eq.dwmetric}) with $e^{A(r)}$ in (\ref{phiA}) and scalars
\be
\label{eq.D2_solu}
z(r)=\z(r) \equiv \frac{1-e^{\varphi(r)}}{1+e^{\varphi(r)}}=\frac{(g \,r)^2 - 2^4}{(g \, r)^2 + 2^4} \; , 
\ee
with $e^{\varphi(r)}$ given also in (\ref{phiA}). In our parameterisation, SO(7)-invariant solutions within the SU(3)-invariant sector are characterised by $\,\textrm{Re}z=\textrm{Re}\z\,$ and $\,\textrm{Im}z=\textrm{Im}\z=0\,$. These conditions are indeed met by (\ref{eq.D2_solu}). Within the SU(3)-invariant sector, this solution preserves four supercharges.

The domain wall (\ref{eq.dwmetric}),  (\ref{phiA}), (\ref{eq.D2_solu}) is the effective four-dimensional description, within $D=4$ $\cN=8$ supergravity with a purely electric ISO(7) gauging \cite{Hull:1984yy}, of the near horizon D2-brane solution (\ref{D2embedding}), (\ref{phiA}) of massless type IIA. The constant $g$ is reinterpreted in this context as the supergravity gauge coupling. This solution is, in turn, dual to three-dimensional $\cN=8$ SYM. The absence  of AdS vacua of the purely electric ISO(7) gauging renders this supergravity inappropriate to study holographically IR superconformal phases of three-dimensional $\cN=8$ SYM. In contrast, the dyonic ISO(7) gauging \cite{Guarino:2015qaa} does possess AdS vacua. In this section we will show that there exist domain wall solutions of dyonic ISO(7) supergravity that interpolate between the effective $D=4$  description  (\ref{eq.D2_solu}) of the D2-brane in the UV and these AdS fixed points in the IR. We will focus on domain walls and IR fixed points with at least SU(3) symmetry, thus contained within the subsector of dyonic ISO(7) supergravity reviewed in section \ref{sec:SU3sector}.

More concretely, we will show the existence of supersymmetric domain wall solutions of the $g \neq 0$, $m \neq 0$ BPS equations (\ref{eq.BPSeqs}), with metric (\ref{eq.dwmetric}) and running scalars $z(r)$, $\zeta_{12}(r)$, that interpolate between a domain wall (\ref{eq.dwmetric}) in the UV supported by scalars
\begin{equation}
\label{eq.D2asymptotics}
\begin{array}{llll}
z & = & \dfrac{(g\,r)^2-2^4 \left(1+\dfrac{m}{g}\, f_1(r)\right)}{(g\,r)^2+2^4 \left(1+\dfrac{m}{g}\, f_1(r)\right)} + i\, \dfrac{m}{g} \dfrac{ f_3(r)}{\big(2^4+(g\,r)^2\big)^2} & , \\[10mm]
\z & = & \dfrac{(g\,r)^2-2^4 \left(1+\dfrac{m}{g}\, f_2(r)\right)}{(g\,r)^2+2^4 \left(1+\dfrac{m}{g}\, f_2(r)\right)} + i\, \dfrac{m}{g} \dfrac{ f_4(r)}{\big(2^4+(g\,r)^2\big)^2} & ,
\end{array}
\end{equation}
and each of the AdS fixed points recorded in table \ref{tab.SU3fps} in the IR. In (\ref{eq.D2asymptotics}), $g$ and $m =gc$ are again the electric and magnetic couplings of dyonic ISO(7) supergravity, and $f_1(r) , \ldots , f_4(r)$ are real functions of the transverse coordinate $\,r\,$ in (\ref{eq.dwmetric}) to be determined shortly. In the limit $c \equiv m/g \rightarrow 0$ with $g$ finite (and even at both $c \equiv m/g$ and $g$ finite, also in the deep UV, see below), the configuration (\ref{eq.D2asymptotics}) reduces to the $D=4$ description (\ref{eq.D2_solu}) of the near-horizon D2-brane solution. Using the SU(3)-invariant consistent uplift formulae  \cite{Varela:2015uca}  of dyonic ISO(7) supergravity into massive type IIA \cite{Guarino:2015jca,Guarino:2015vca}, the configuration (\ref{eq.dwmetric}), (\ref{eq.D2asymptotics}) uplifts to the most general deformation of the near horizon D2-brane solution (\ref{D2embedding}) that is small in the Romans mass $\hat F_\0 = m$ and preserves at least the SU(3) subgroup of SO(7). By the equality between (quantised) Romans mass and three-dimensional Chern-Simons coupling, $\hat F_\0 = k/(2\pi \ell_s)$ \cite{Gaiotto:2009mv,Guarino:2015jca}, this massive type IIA configuration is dual to the most general deformation of $\cN=8$ SYM with Chern-Simons terms and adjoint matter with at least SU(3) flavour.

An uneventful integration shows that the configuration (\ref{eq.D2asymptotics}) solves the BPS equations (\ref{eq.BPSeqs}) at linear order in $\,c=m/g\,$ provided the functions $f_1(r) , \ldots , f_4(r)$ are given by
\begin{equation}
\label{eq.f's_D2}
\begin{array}{llllllll}
f_1 & = & \dfrac{c_1}{g\, r} + \dfrac{c_2}{(g\,r)^8} & \hspace{5mm} , \hspace{5mm} & \hspace{5mm} f_3 & = & \dfrac{c_3}{(g\,r)^4} + \dfrac{c_4}{(g\,r)^{12}}  - \dfrac{196608}{7(g\,r)^4} &  , \\[5mm]
f_2 & = & \dfrac{c_1}{g\, r} - \dfrac{3}{4} \, \dfrac{c_2}{(g\,r)^8} & \hspace{5mm} , \hspace{5mm} & \hspace{5mm}f_4 & = &  - \dfrac{3}{4} \, \dfrac{c_3}{(g\,r)^4} +  \dfrac{c_4}{(g\,r)^{12}} - \dfrac{196608}{7(g\,r)^4} & ,
\end{array}
\end{equation}
for arbitrary real integration constants $c_1 , \ldots , c_4$. As (\ref{eq.f's_D2}) shows, all the corrections (with at least SU(3)-invariance) in (\ref{eq.D2asymptotics}) to the effective $D=4$ description (\ref{eq.D2_solu}) of the D2-brane near horizon created by the dyonic parameter $c=m/g$ are suppressed by inverse powers of $g \, r$. Thus, the configuration (\ref{eq.D2asymptotics}), (\ref{eq.f's_D2}) can indeed serve as the asymptotic UV of domain walls with at least SU(3) symmetry, and only as the UV, not as the IR.  This indicates that the addition of the Romans mass or, holographically, of Chern-Simons-matter with at least SU(3) flavour, is a relevant deformation of the $\cN=8$ SYM UV action.   In the deep UV, $r \rightarrow + \infty$, the scalars (\ref{eq.D2asymptotics}), (\ref{eq.f's_D2}) approach the boundary of their Poincar\'e disks, $z \rightarrow 1$, $\z \rightarrow 1$, through the asymptotic D2-brane behaviour  (\ref{eq.D2_solu}). Note also from (\ref{eq.f's_D2}) that  it can never happen that $\,f_3=f_4=0\,$. This implies that the deformation by the Romans mass always breaks the SO(7) R-symmetry of the D2-brane solution (\ref{eq.D2_solu}), as expected.

It is sufficient for our purposes to consider (\ref{eq.D2asymptotics}), (\ref{eq.f's_D2}) with $\,c= m/g\,$ small as the UV behaviour of our SYM to CFT flows. The reason for this is that we want to treat the Romans mass as a perturbation of the D2-brane configuration (\ref{eq.D2_solu}). As $\,m\,$ becomes large compared to $\,g\,$, (\ref{eq.D2_solu}) should be ultimately replaced with the configuration
\be
\label{eq.g=0}
z(r)=\z(r)=\frac{ (7\,m\,r)^{2/7} - 2^{4/7}}{  (7\,m\,r)^{2/7} + 2^{4/7} }
\hspace{10mm} , \hspace{10mm}
e^{A(r)} =  (7 \, m \, r)^{1/7} \ ,
\ee
which solves (\ref{eq.BPSeqs}) with $g =0$, $m \neq 0$. This is no longer a solution of $\cN=8$ supergravity with a dyonic $\textrm{ISO}(7)=\textrm{CSO}(7,0,1) $ gauging. Instead, it is a solution of $\cN=8$ supergravity with a purely magnetic nilpotent CSO$(1,0,7)$ gauging. As argued briefly at the end of section 2.3 of \cite{Guarino:2015qaa}, the latter gauging uplifts to massive type IIA on $T^6$ rather than $S^6$. As a result, (\ref{eq.g=0}) gives rise to a ten-dimensional solution, presumably related to the D8-brane, very different from a near horizon geometry like (\ref{D2embedding}). We will not discuss this configuration any further.

\subsection{Flow into the $\cN=1$ G$_2$ conformal phase} \label{subsec:G2flows}

For the sake of stability of our numerical routine, we integrate the equations (\ref{eq.BPSeqs}) with a shooting method that starts near each of the IR fixed points. Let us first look at supersymmetric flows with IR end in the $\textrm{G}_{2}$ fixed point. By inspection of table~\ref{tab.deltas} we conclude that it is indeed possible for this point to serve as the IR ($r \to -\infty$) endpoint of a domain wall. The reason is that there is an irrelevant mode with $\tilde \Delta= \Delta_- = 1-\sqrt{6}<0$. This mode is, moreover, non-normalisable, in agreement with the expectation that the dual flow should be triggered by deformations of the field theory Lagrangian. Perturbing the IR CFT by this mode corresponds to  adding a scale to the IR CFT. Since this is the only scale in the theory, all of its values are equivalent and there is only one physically independent RG flow. This has a counterpart in the numerical integration: the $a_i$ and $\alpha_i$ coefficients in \eqref{eq.perturbAdS} for this flow turn out to depend on only one parameter. This can be fixed by a shift of the transverse coordinate. Although the actual value of this parameter is immaterial, one must pick the correct sign, see appendix~\ref{App:boundary_conds}. 

\begin{figure}[t!]
\begin{center}
\includegraphics[width=0.4\textwidth]{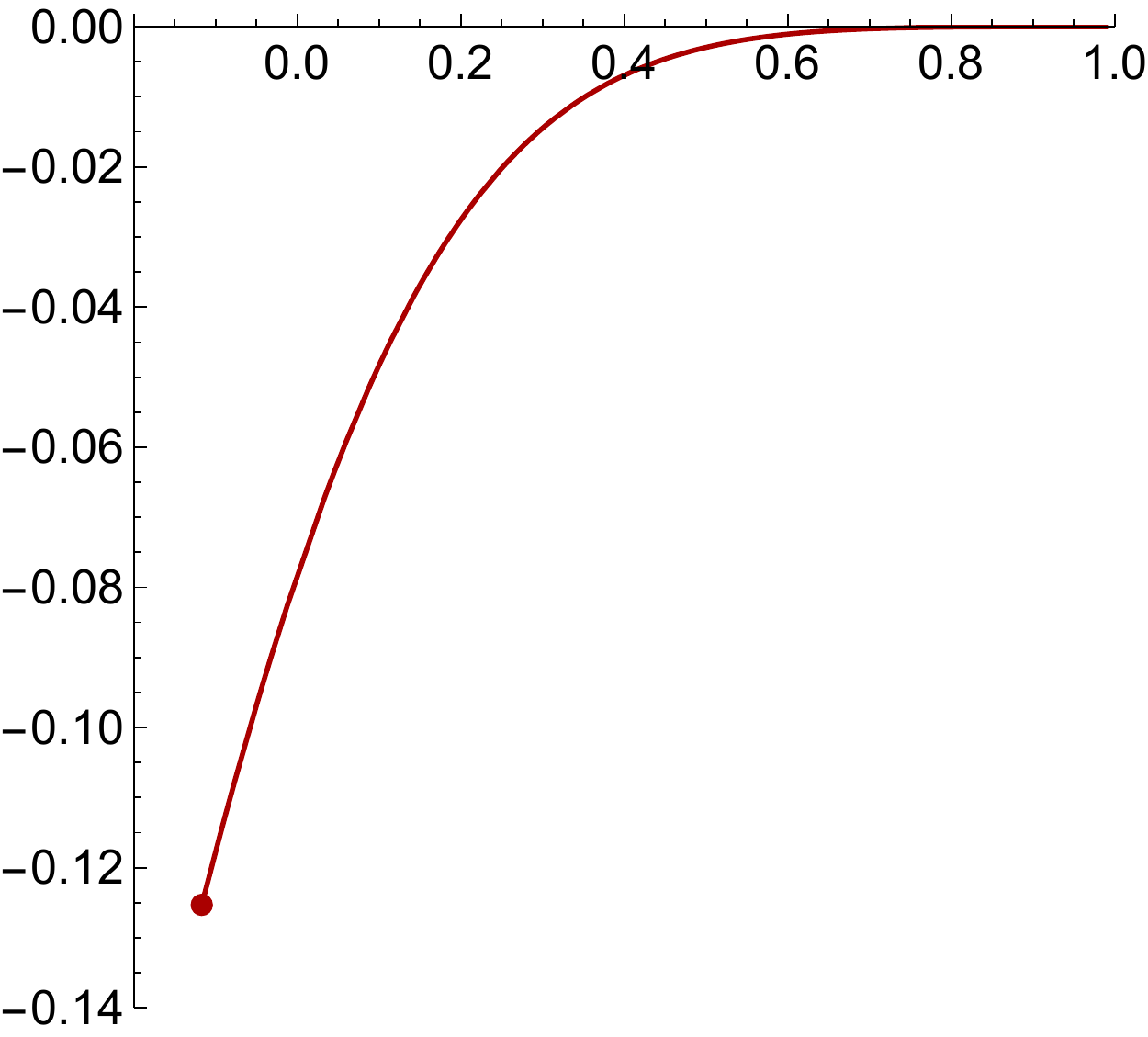}
\put(-205,90){\large Im$z$}
\put(10,150){\large Re$z$}
\put(-140,10){\small G$_2$}
\caption{Trajectory of the SYM to CFT supersymmetric domain wall solution with the $\cN=1$ G$_2$ fixed point (red dot) in the IR. $\textrm{G}_{2}$-invariance requires $\,z=\z\,$ along the flow. This and all other plots in the paper have been generated with $g=m=1$, {\it i.e.}, $c=1$.}
\label{fig.SYMtoG2domainwall}
\end{center}
\end{figure}

As a result, our shooting method produces a unique $\cN=1$ flow with IR endpoint in the $\cN=1$ G$_2$ fixed point, which has $\,z=\z\,$ all along the flow. This condition ensures that the entire flow is G$_2$-symmetric, like the IR fixed point. In the UV, the flow asymptotes to (\ref{eq.D2asymptotics}), (\ref{eq.f's_D2}) with $\,c_2 = c_3 = 0\,$. In other words, the UV end of this flow is dominated by the effective $D=4$ description (\ref{eq.D2_solu}) of the near horizon D2-brane, with subleading corrections as in (\ref{eq.D2asymptotics}), (\ref{eq.f's_D2}) produced by the non-zero $\,c=m/g\,$. The fact that  $\,c_2 = c_3 = 0\,$ ensures that this deformation by the Romans mass is G$_2$-symmetric.  In field theory terms, the UV dynamics is governed by the SYM theory living in the worldvolume of the D2-branes, since the YM term of the action is irrelevant and dominates over the Chern-Simons term at high energies. At a certain energy scale the gluons of this theory become massive and decouple at low energies, where the theory crossovers to a Chern-Simons-matter theory and becomes conformal in the IR. The numerical trajectory of this flow in the $\,z=\z\,$ Poincar\'e disk is depicted in figure~\ref{fig.SYMtoG2domainwall}. Note that the plot approaches $\,z=\z \to 1\,$ in the deep UV, $r\to +\infty$, in agreement with (\ref{eq.D2asymptotics}), (\ref{eq.f's_D2}), and (\ref{eq.D2_solu}).

The consistent uplifting formulae of dyonic ISO(7) supergravity into massive type IIA \cite{Guarino:2015jca,Guarino:2015vca} can be used to write the ten-dimensional solution corresponding to this domain wall.  Using the G$_2$-invariant restriction of these formulae given in (4.3), (4.4) of \cite{Guarino:2015vca}, the result is
\begin{equation}
\label{G2embeddingGeom}
\hspace{-3.4mm}
\begin{array}{rll}
d \hat{s}_{10}^2 &=& e^{\frac{3}{4} \varphi(r)} \, X(r)^{\frac{3}{4}} \, \big(e^{2A(r)} \, \eta_{\alpha\beta} \, \d x^{\alpha} \, \d x^{\beta} + \d r^2 \big) +  g^{-2} \, e^{-\frac{1}{4} \varphi(r)} \, X(r)^{-\frac{1}{4}} \,  ds^2(S^6) \ , \\[5pt]
e^{\hat \phi} &=& e^{\frac{5}{2} \varphi(r)} \, X(r)^{-\frac32}  \ , \\[5pt]
\hat F_{(4)} &=&  \left[ g \, e^{\varphi(r)} \, X(r)^2 \,  \big( 5 - 7 \, e^{2\varphi(r)} \chi(r)^2  \big) + m \,e^{7\varphi(r)} \chi(r)^3 \right]  \vol_{4}  \\[5pt]
&& \, + \,  g^{-3} \,  \d \chi(r)  \wedge \textrm{Im} \,  \Upomega     +   \left[ \tfrac{1}{2} \, m \, g^{-4} \,  e^{4 \varphi(r)}  \, \chi(r)^2  \, X(r)^{-2}  -2 \, g^{-3} \,  \chi(r)  \right] {\cal J} \wedge {\cal J}  \ ,  \\[5pt]
\hat H_{(3)} &=& g^{-2}  \, \d \Big(  e^{2 \varphi(r)}  \, \chi(r)  \, X(r)^{-1} \Big) \wedge {\cal J} + 3 \, g^{-2} \,   e^{2 \varphi(r)}  \, \chi(r)  \, X(r)^{-1}  \, \textrm{Re} \, \Upomega \ ,  \\[5pt]
\hat F_{(2)} &=& m \, g^{-2} \,   e^{2 \varphi(r)}  \, \chi(r)  \, X(r)^{-1}  \, {\cal J} \ ,
\end{array}
\end{equation} 
together with the general relation $\hat F_\0 = m$. The external volume form is given by $\,\vol_{4}= \frac{1}{3!} \, e^{3A(r)} \, \epsilon_{\alpha \beta \gamma} \, \d x^\alpha \wedge \d x^\beta \wedge \d x^\gamma \wedge \d r \,$, $\,{\cal J}\,$ and $\,\Upomega\,$ are the G$_2$-invariant nearly-K\"ahler forms on the six sphere, regarded as the homogeneous space $\,S^6 = \textrm{G}_2/\textrm{SU}(3)\,$, and $ds^2(S^6)$ the Einstein metric they determine. The latter coincides with the usual, round Einstein metric that appears in (\ref{D2embedding}). Also, $\,X(r) \equiv 1+e^{2 \varphi(r)} \chi(r)^2 \,$, and the transverse functions $\,\varphi(r)\,$ and $\chi(r) \,$ are given in terms of the numerical $z(r)$ in figure \ref{fig.SYMtoG2domainwall} by
\begin{equation}
e^{\varphi(r)}=  2 \, \left( \frac{\textrm{Re}z(r)-1}{|z(r)|^2-1} \right) -1
\hspace{6mm} , \hspace{6mm}
\chi(r)= 2 \, \frac{\textrm{Im}z(r)}{(\textrm{Im} \, z(r))^2+(\textrm{Re}z(r)-1)^2} \ .
\end{equation}

\subsection{Flows into the $\cN=2$ $\textrm{SU}(3) \times \textrm{U}(1)$ conformal phase} \label{subsec:SU3U1flows}

We have found similar supersymmetric flows that drive the UV asymptotic D2-brane configuration   (\ref{eq.D2asymptotics}), (\ref{eq.f's_D2}) into the IR conformal phases with $\cN=2$ $\textrm{SU}(3) \times \textrm{U}(1) $ and $\cN=1$ SU(3) symmetries. In contrast to the unique G$_2$-invariant flow, RG flows with IR endpoints in either of these two phases come in one-parameter families.

Let us first focus on RG flows with IR endpoint on the $\cN=2$ $\textrm{SU}(3) \times \textrm{U}(1)$ conformal phase. Within the SU(3)-invariant truncation of dyonic ISO(7) supergravity that we are considering, this fixed point displays two irrelevant and non-normalisable modes. These are the negative entries $\tilde \Delta = \Delta_- =  (1-\sqrt{17})/2$ and $\tilde \Delta = \Delta_- =  (3-\sqrt{17})/2$ marked in red in table \ref{tab.deltas}. The existence of these two modes in principle allows for a two-parameter family of flows but, as discussed in more detail in appendix~\ref{App:boundary_conds}, one of these parameters can be fixed (up to a sign) by a shift of the domain wall transverse coordinate $r$. The one-parameter freedom of this family of domain walls is reflected in the field theory side: the addition of an irrelevant deformation to a CFT adds a scale, $E_1$,  in the field theory that modifies the UV. If a second scale, $E_2$, is further added, they cannot be both removed simultaneously. As a result, there is a one-parameter family of inequivalent physics parameterised by the ratio $E_2/E_1$.

\begin{figure}[t]
\begin{center}
\hspace{6mm}
\includegraphics[width=0.4\textwidth]{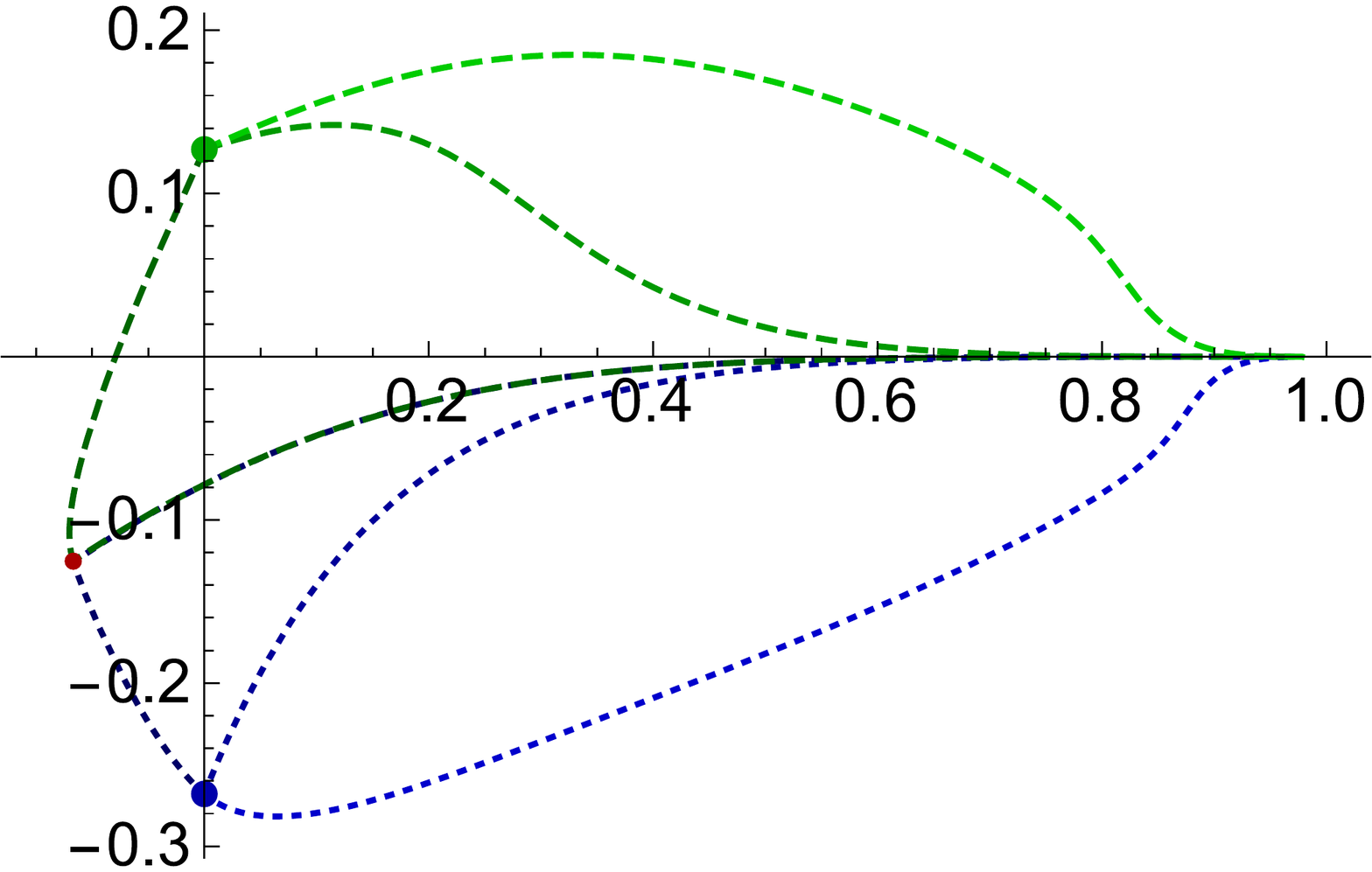}\hspace{2cm}
\includegraphics[width=0.4\textwidth]{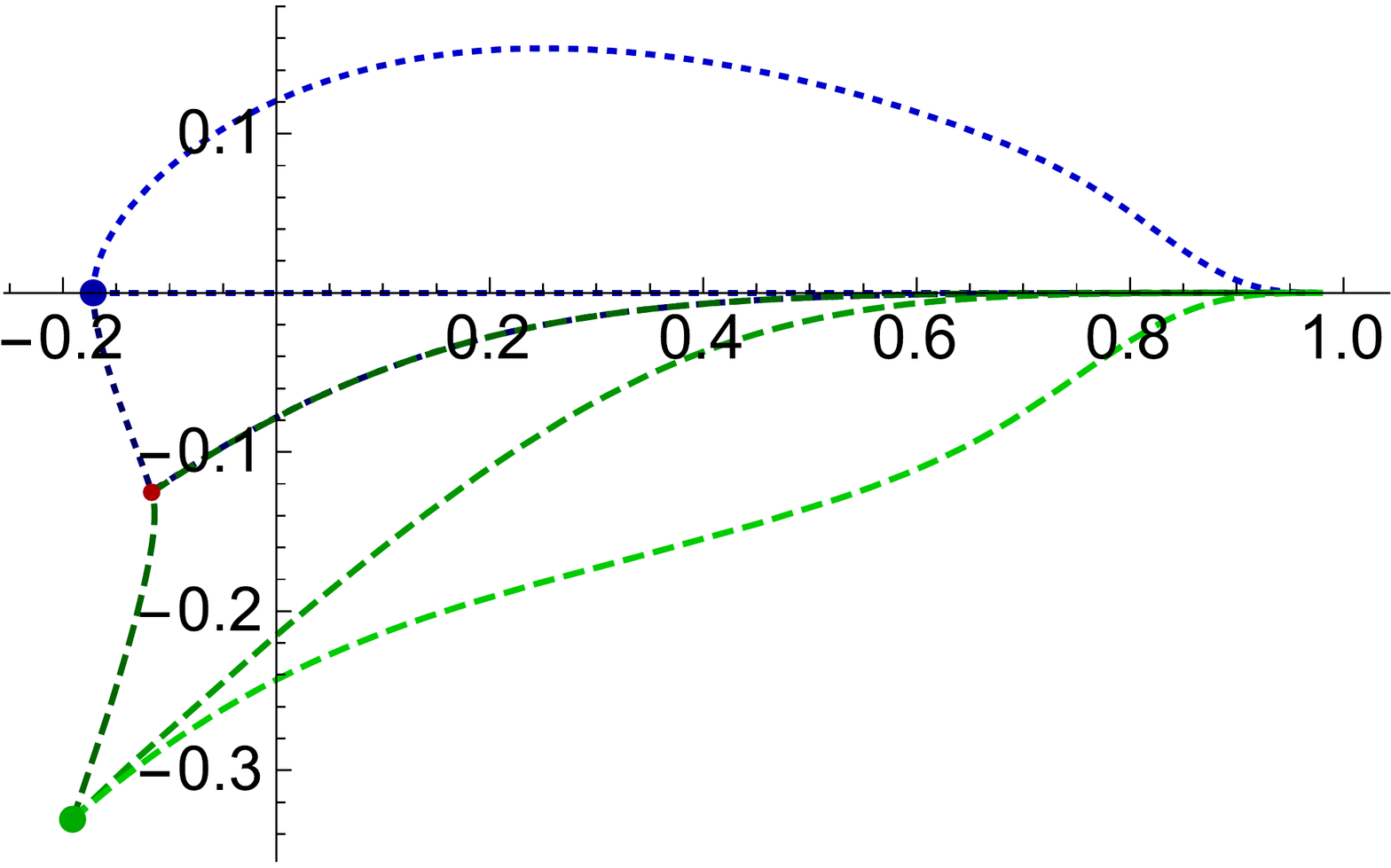}
\put(-138,180){\large Im$\z$}
\put(-25,110){\large Re$\z$}
\put(-378,180){\large Im$z$}
\put(-256,105){\large Re$z$}
\put(-415,100){\small G$_2$}
\put(-175,110){\small G$_2$}
\put(-218,145){\small SU(3)$\times$U(1)}
\put(-380,55){\small SU(3)$\times$U(1)}
\put(-190,60){\small SU(3)}
\put(-420,155){\small SU(3)}
\vspace{-15mm}
\caption{Numerically integrated trajectories of some SYM to CFT flows. Three flows in each one-parameter family with IR dominated by the $\cN=2$ SU(3)$\times$U(1) and the $\cN=1$ SU(3) fixed points are shown with dotted blue lines and dashed green lines, respectively. For each family it is shown a generic flow, the `direct' flow and a flow that runs very close to the common boundary set by the G$_2$ flow of figure \ref{fig.SYMtoG2domainwall} (not shown here). The latter flows appear almost superimposed along this common boundary.}\label{fig.cones}
\end{center}
\end{figure}

All the solutions in the family preserve at least SU(3) symmetry and $\cN=1$ supersymmetry along the flow, with an enhancement to $\textrm{SU}(3) \times \textrm{U}(1)$ symmetry and $\,\cN=2\,$ supersymmetry at the IR endpoint. The family is bounded by the G$_2$-invariant flow of section \ref{subsec:G2flows}, and a flow that connects the $\cN=1$ G$_2$ fixed point in the UV with the $\cN=2$ $\textrm{SU}(3) \times \textrm{U}(1) $ point in the IR. The latter will be further discussed in section \ref{CFTtoCFT}. The superconformal field theory dual to the $\cN=2$ $\textrm{SU}(3) \times \textrm{U}(1)$ IR fixed point is in the class of theories considered in \cite{Schwarz:2004yj}. As explained in \cite{Guarino:2015jca}, this corresponds to $\cN=2$ Chern-Simons theory with a single gauge group SU$(N)$, coupled to three adjoint chiral fields in the fundamental of the SU(3) flavour group and subject to a cubic superpotential. The free energy of this field theory on $S^3$ was computed  \cite{Guarino:2015jca} using localisation \cite{Kapustin:2009kz,Jafferis:2010un,Hama:2010av} and shown to match the gravitational free theory of the dual AdS$_4$ massive type IIA configuration.

One flow in the family is special in that it minimises the trajectory in the $(z, \z)$ scalar space between the deep UV D2-brane solution (the boundary $z \rightarrow 1$, $\z \rightarrow 1$  of the Poincar\'e disks) and the $\cN=2$ $\textrm{SU}(3) \times \textrm{U}(1)$ IR endpoint. The bosonic symmetry of this `direct' flow is enhanced to the full $\textrm{SU}(3) \times \textrm{U}(1)$ symmetry of the IR fixed point. Also the supersymmetry along this flow is enhanced, to $\cN=2$ (four Poincar\'e supercharges, see the comments on page \pageref{Ncomments}). This `direct' flow occurs with $\bar{\zeta}_{12} = \z$. This condition is in fact responsible for the supersymmetry and U(1) symmetry enhancement: the two \mbox{SU(3)-invariant} eigenvalues of the $\cN=8$ gravitino mass matrix, namely, the two superpotentials discussed in section \ref{subsec:Floweqs}, become degenerate when $\bar{\zeta}_{12} = \z$. Gauged $\cN=2$ supergravity techniques similar to those used in \cite{Tarrio:2013qga} allow us to determine the analytic trajectory in scalar space  of the $\textrm{SU}(3) \times \textrm{U}(1)$-invariant `direct' flow. In the the $(z, \z)$ Poincar\'e disk parameterisation that we are using, this trajectory is given by\footnote{In the parameterisation used in section 3.1 of \cite{Guarino:2015qaa}, equations (\ref{eq:N=2flow}) read  $g \, t  \, \bar t  \, ( t + \bar t) -m  =0$ and $ \zeta = \tilde{\zeta} = 0 $.}
\begin{eqnarray} \label{eq:N=2flow}
2ig (1+z)(1+\bar z) (z-\bar z) - m \,  (1-z)^2 (1- \bar z)^2 = 0 \; , \qquad \quad  \bar{\zeta}_{12} = \z \; .
\end{eqnarray}
We have numerically verified the validity of these equations. They should be treasured, as analytic results in the holographic RG flow literature do not abound.

The numerically integrated trajectories of three flows in this one-parameter family are depicted with dotted blue lines in figure ~\ref{fig.cones}. The trajectory farthest from the horizontal axes corresponds to a generic flow. The middle trajectory is that, (\ref{eq:N=2flow}), of the `direct' flow. A third flow is depicted with trajectory very close to the boundary of the family. This flow leaves the deep UV D2-brane solution, $z \rightarrow 1$, $\z \rightarrow 1$, following closely the G$_2$-invariant flow of section \ref{subsec:G2flows}. It displays walking dynamics dominated by the G$_2$ fixed point for a long parametric time or, holographically, for a long range of the ratio of scales $E_2/E_1$. This flow eventually approaches the $\cN=2$ $\textrm{SU}(3) \times \textrm{U}(1)$ fixed point in the IR following closely the G$_2$ to $\textrm{SU}(3) \times \textrm{U}(1)$ flow of section \ref{CFTtoCFT}. 

Finally, we note that the entire one-parameter family of flows can be uplifted to massive type IIA using the consistent truncation formulae of \cite{Guarino:2015jca,Guarino:2015vca} particularised to the SU(3)-invariant sector \cite{Varela:2015uca}.

\subsection{Flows into the $\cN=1$ $\textrm{SU}(3)$ conformal phase} \label{subsec:SU3flows}

A very similar story arises for supersymmetric domain walls that land on the $\cN=1$ $\textrm{SU}(3)$ fixed point. Holographic RG flows into this point are also driven by two irrelevant and non-normalisable modes, marked in red in table \ref{tab.deltas}, with $\,\tilde \Delta = \Delta_- <0\,$. This leaves, upon gauge fixing of the transverse coordinate as in the previous case, a one-parameter family of $\cN=1$ SU(3)-invariant flows that interpolate between  the D2-brane behaviour (\ref{eq.D2asymptotics}) in the UV and the $\cN=1$ SU(3) fixed point in the IR. This family is bounded by the G$_2$-invariant flow of section \ref{subsec:G2flows} and the flow discussed in section \ref{CFTtoCFT} that connects the G$_2$ fixed point in the UV and SU(3) point in the IR. One of the flows in the family has minimal trajectory in scalar space, but no symmetry or supersymmetry enhancements occur in this case. Finally,  this family of flows uplifts on $S^6$ to massive type IIA using the SU(3)-invariant specialisation \cite{Varela:2015uca} of the reduction formulae of \cite{Guarino:2015jca,Guarino:2015vca}. The IR fixed point corresponds to the $\cN=1$ SU(3)-invariant AdS$_4$ solution of massive IIA found in \cite{Varela:2015uca}. 

The numerical trajectories of three flows in this family are plotted with dashed green lines in figure~\ref{fig.cones}. These correspond to a generic flow, to the `direct' flow with minimal trajectory, and to a flow that follows closely the boundary and displays walking dynamics governed by the G$_2$ fixed point.

%%%%%%%%%%%%%%%%%
%%%%%%%%%%%%%%%%%

\section{CFT to CFT flows} \label{CFTtoCFT}

%%%%%%%%%%%%%%%%%
%%%%%%%%%%%%%%%%%

We now turn to discuss RG flows that interpolate between superconformal phases with at least SU(3) flavour symmetry at both UV and IR endpoints. An argument based on the following hierarchy of cosmological constants 
\be
\label{eq.potentialhierarchy}
0 \, > \, V_{*}^{\textrm{G}_2} \,>\, V_{*}^{\textrm{SU}(3)\times \textrm{U}(1)} \,>\, V_{*}^{\textrm{SU}(3)} \ 
\ee
(see table \ref{tab.SU3fps}), suggests that there may be at most three types of flows of this type: flows that originate at the G$_2$ point whose IR is dominated by either the $\textrm{SU}(3) \times \textrm{U}(1)$ or the SU(3) point, and flows that originate at the  $\textrm{SU}(3) \times \textrm{U}(1)$ phase and reach the SU(3) point in the IR. We find that only the first two types of flows exist and are, moreover, unique within the SU(3)-invariant sector. These two flows have already been noted in sections \ref{subsec:SU3U1flows} and \ref{subsec:SU3flows}. The latter type of RG flows is not realised.

For these RG flows to exist, the supersymmetric CFT with G$_2$ symmetry must possess at least two relevant, $\Delta_+ <3$, deformations. Each of these would be responsible to trigger flows into either  IR phase, SU(3) and $\textrm{SU}(3) \times \textrm{U}(1)$. There are indeed two such modes, that we called $1$ and $2$ in table~\ref{tab.deltas}. For both of them, the non-normalisable fall-off, $\tilde{\Delta} = \Delta_->0$ happens to be activated in the perturbative solution \eqref{eq.perturbAdS} about the UV G$_2$ point. The (normalisable) mode that falls-off with  $\,\Delta_3=1+\sqrt{6}\,$ has to be switched off, as it corresponds to an irrelevant operator. In summary,  we expect that there are two independent linear combinations of operators with dimensions $\Delta_\pm =2\pm1/\sqrt{6}\,$: one that makes the theory flow to the $\textrm{SU}(3) \times \textrm{U}(1)$ fixed point and the other to the $\textrm{SU}(3)$ point.

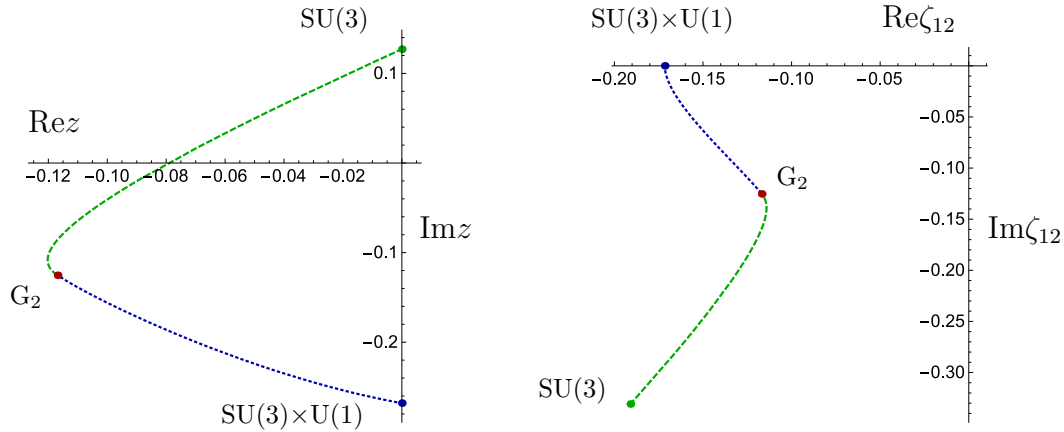
\begin{figure}[t]
\begin{center}
\includegraphics[width=0.35\textwidth]{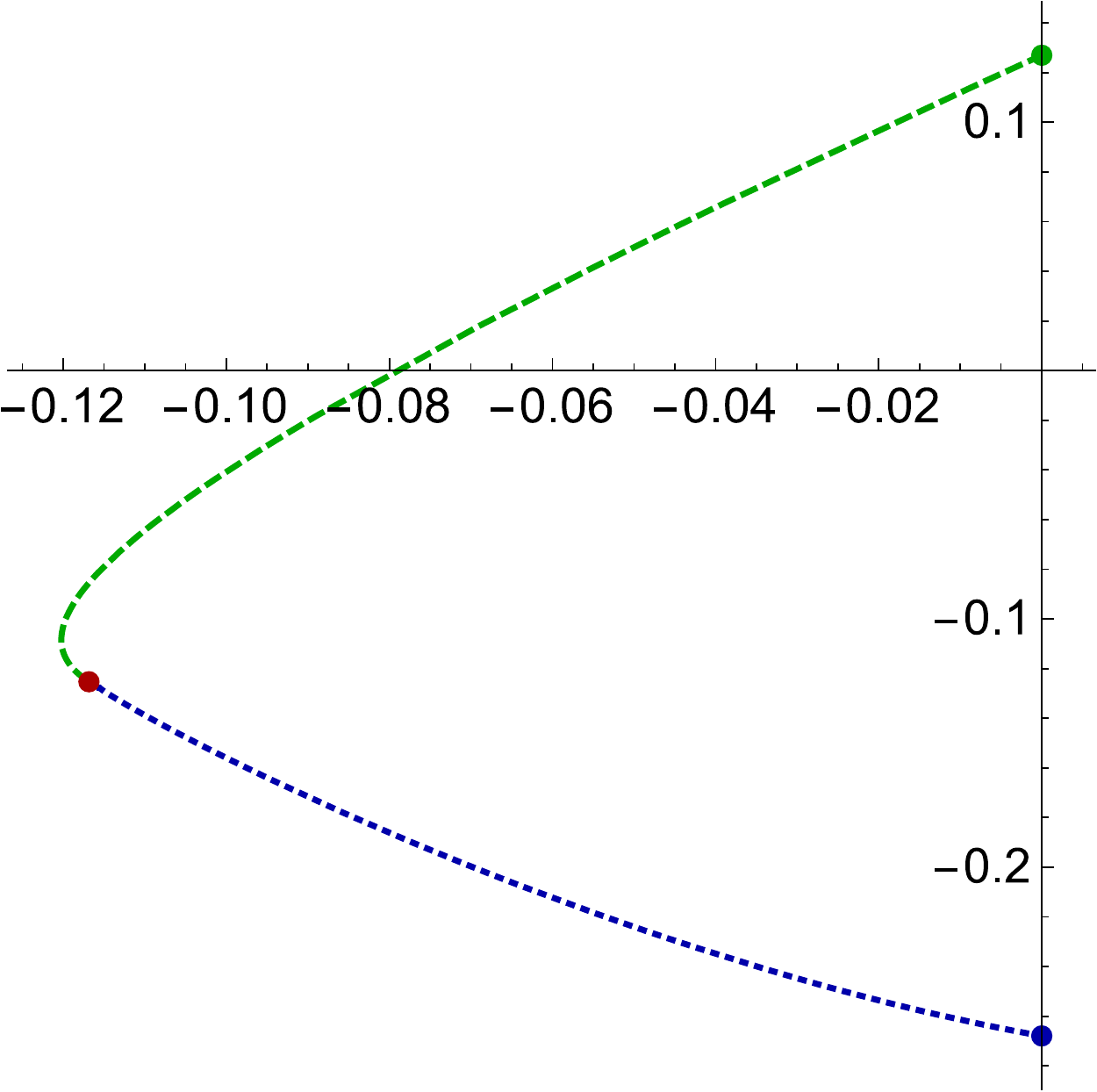}
\hspace{20mm}
\includegraphics[width=0.35\textwidth]{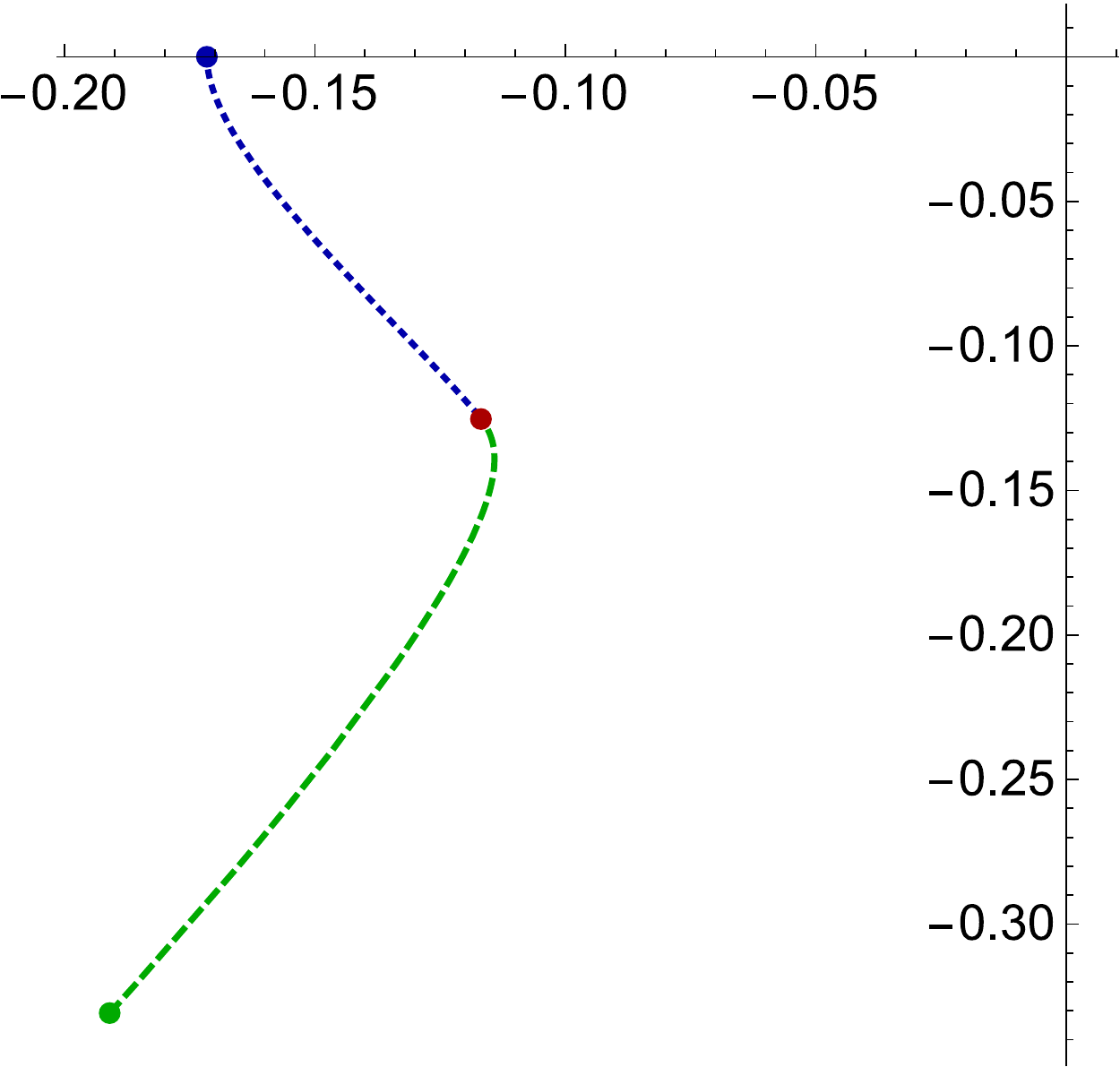}
\put(-40,150){\large Re$\z$}
\put(0,70){\large Im$\z$}
\put(-215,70){\large Im$z$}
\put(-363,110){\large Re$z$}
\put(-370,45){\small G$_2$}
\put(-80,90){\small G$_2$}
\put(-150,150){\small SU(3)$\times$U(1)}
\put(-290,0){\small SU(3)$\times$U(1)}
\put(-170,10){\small SU(3)}
\put(-260,150){\small SU(3)}
\caption{Numerically integrated trajectories of the CFT to CFT flows that interpolate between the G$_2$ fixed point (red dot) in the UV and the SU(3)$\times$U(1) (dotted blue line) or the SU(3) (dashed green line) fixed point in the IR.}
\label{fig.AdSdomwalls}
\end{center}
\end{figure}

A numerical integration of the BPS equations (\ref{eq.BPSeqs}) shows that both such BPS domain walls do indeed exist. As in the previous section, it is simpler to integrate the equations starting near the $\textrm{SU}(3) \times \textrm{U}(1)$ or SU(3) IR fixed points. For generic IR boundary conditions compatible with turning on irrelevant deformations, one recovers the two families of flows with UV dominated by the D2-brane solution that we discussed in the previous section. However, for both IR fixed points we find that a fine-tuning of the boundary conditions changes the UV behaviour. In both cases, the UV of these fine-tuned domain walls is dominated by the $\cN=1$ G$_2$-invariant phase. Moreover, the numerical integration shows that precisely the expected modes that we described above drive these flows into the UV G$_2$ phase. Both flows are $\cN=1$ and SU(3)-invariant. Finally, each of them serves as a boundary of the one-parameter family of flows discussed in section \ref{sec:SYMtoCFT}
with the same IR fixed point.

Figure~\ref{fig.AdSdomwalls} depicts the numerically integrated trajectories of these CFT to CFT flows in scalar space.

\section{Discussion} \label{sec:Discussion}

We have studied supersymmetric domain wall solutions of $D=4$ $\cN=8$ dyonically-gauged ISO(7) supergravity that display at least SU(3) invariance. The domain walls whose UV is dominated by (the four-dimensional description of) the D2-brane, (\ref{eq.dwmetric}),  (\ref{phiA}), (\ref{eq.D2_solu}), have a clear AdS/CFT interpretation. They describe holographically the RG evolution of three-dimensional $\cN=8$ SYM theory when the latter is  augmented with Chern-Simons terms and adjoint supersymmetric matter with at least SU(3) flavour symmetry. We have also constructed domain walls between the conformal phases of the D2-brane with at least SU(3) symmetry.

Our analysis uncovers a rich web of supersymmetric flows between $\cN=8$ SYM and different IR superconformal phases. In the supergravity sector under investigation, there exists a unique domain wall with G$_2$ symmetry along the entire solution that drives the D2-brane worldvolume theory into the $\cN=1$ phase with that flavour symmetry.  This unique $\cN= 1$ G$_2$-invariant flow lies at the common boundary of two one-parameter families of $\cN=1$ SU(3)-invariant flows. One family contains flows that interpolate between the D2-brane behaviour in the UV and the $\cN=2$ $\textrm{SU}(3) \times \textrm{U}(1)$-invariant fixed point in the IR. One of the flows in this family has its supersymmetry and symmetry enhanced to $\cN=2$ and $\textrm{SU}(3) \times \textrm{U}(1)$. The other one-parameter family of flows interpolates between the UV D2-brane worldvolume theory and an $\cN=1$ SU(3)-invariant phase in the IR. In addition, these families are also bounded by unique $\cN=1$ SU(3)-invariant flows with UV origin in the G$_2$ phase and IR endpoint in either the $\cN=2$ $\textrm{SU}(3) \times \textrm{U}(1)$ or the $\cN=1$ SU(3) points. See figures \ref{fig.SYMtoG2domainwall},  \ref{fig.cones} and \ref{fig.AdSdomwalls} for a graphical summary of this web of supersymmetric RG flows. Our results are numerical and, for the flow with enhanced symmetry into the $\cN=2$ point, also analytic: see equation (\ref{eq:N=2flow}).

To some extent, these results parallel  the situation for flows of the M2-brane field theory with at least SU(3) symmetry. These have been studied \cite{Ahn:2000aq,Ahn:2000mf,Corrado:2001nv,Bobev:2009ms} within the effective description provided by $D=4$ $\cN=8$ supergravity with a (purely electric) SO(8) gauging  \cite{deWit:1982ig}. This gauged supergravity also has critical points with $\cN=1$ G$_2$-symmetry and  $\cN=2$ $\textrm{SU}(3) \times \textrm{U}(1)$ symmetry (but not with  $\cN=1$ SU(3) symmetry). These critical points give rise, by the consistency of the truncation of $D=11$ supergravity on $S^7$ \cite{deWit:1986iy}, to AdS$_4$ solutions in $D=11$. A one-parameter family of $\cN=1$ SU(3)-invariant flows exists that drives the M2-brane field theory into the  $\cN=2$ $\textrm{SU}(3) \times \textrm{U}(1)$ conformal phase \cite{Bobev:2009ms}. This family contains the `direct' flow of \cite{Ahn:2000aq,Benna:2008zy} and is bounded by the flow into the $\cN=1$ G$_2$ phase \cite{Ahn:2000mf} and by the flow from the G$_2$ phase in the UV and the $\cN=2$ $\textrm{SU}(3) \times \textrm{U}(1)$ point in the IR \cite{Bobev:2009ms}. With the exception of the latter, the UV of these flows is dominated by the $\cN=8$ SO(8) point of the SO(8) gauging, dual to the superconformal field theory of  \cite{Aharony:2008ug}. 

Apart from this important difference (the UV of our flows is generically dominated by the non-conformal $\cN=8$ SYM theory in three dimensions) the situation is quite similar in our present D2-brane case. Another difference between the M2 and D2-brane cases is the presence in the latter of an $\cN=1$ SU(3)-invariant conformal phase with no counterpart in the former. Unlike the purely electric SO(8) gauging  \cite{deWit:1982ig} of $D=4$ $\cN=8$ supergravity, but like the dyonic ISO(7) gauging \cite{Guarino:2015qaa}, the dyonic SO(8) gauging \cite{Dall'Agata:2012bb} has an $\cN=1$ SU(3)-invariant critical point \cite{Borghese:2012zs}. In the dyonic SO(8) gauging, the $\cN=2$  $\textrm{SU}(3) \times \textrm{U}(1)$ point still serves as the IR endpoint of a family of domain walls with UV origin in the $\cN=8$ SO(8) point \cite{Guarino:2013gsa,Tarrio:2013qga}. In addition, the new $\cN=1$ SU(3) point also dominates the IR of a one-parameter family of domain walls. Both families are, also in this case, bounded by the G$_2$-invariant flow and by flows between the G$_2$, $\textrm{SU}(3) \times \textrm{U}(1)$ and SU(3) points \cite{Guarino:2013gsa,Tarrio:2013qga}. The web of domain walls with at least SU(3) invariance in the dyonic ISO(7) gauging is thus much more similar to the web in the dyonic SO(8) gauging \cite{Guarino:2013gsa,Tarrio:2013qga} than to the web in the purely electric SO(8) gauging  \cite{Bobev:2009ms}, also with the same caveat about the generic UV behaviour. No-go statements \cite{deWit:2013ija,Lee:2015xga} against a conventional higher-dimensional origin of dyonic SO(8) supergravity \cite{Dall'Agata:2012bb} seem to preclude a holographic interpretation of the domain walls of \cite{Guarino:2013gsa,Tarrio:2013qga} beyond the strict $N \rightarrow \infty$ limit. This is unlike the domain walls \cite{Ahn:2000aq,Ahn:2000mf,Corrado:2001nv,Bobev:2009ms} of SO(8) electrically gauged supergravity and unlike the domain walls that we have constructed in this paper.

The consistency of the truncation of massive type IIA supergravity on the six-sphere \cite{Guarino:2015jca,Guarino:2015vca} down to dyonic ISO(7) supergravity ensures the existence of a ten-dimensional counterpart to our web of four-dimensional BPS domain walls. In other words, there exists a network of supersymmetric ten-dimensional solutions that interpolate between the AdS$_4$ solutions of massive type IIA with at least SU(3) internal symmetry that were recently constructed in \cite{Guarino:2015jca,Varela:2015uca} (and, in the case of the the $\cN=1$ G$_2$-invariant solution, earlier in \cite{Behrndt:2004km}). We have explicitly written down one of these solutions in equation (\ref{G2embeddingGeom}). More generally, these ten-dimensional solutions can be obtained from the four-dimensional domain walls that we have constructed in this paper using the SU(3)-invariant particularisation  \cite{Varela:2015uca} of the generic $\cN=8$  consistent truncation formulae of massive IIA supergravity on $S^6$  \cite{Guarino:2015jca,Guarino:2015vca}. Previous work on domain walls of the CSO$(p,q,r)$ gaugings and their possible higher-dimensional origin includes \cite{Bergshoeff:2004nq}. See \cite{Haack:2009jg} for a supersymmetry analysis of domain walls already in ten dimensions.

In this paper we have focused on the effect of the Romans mass on the worldvolume theory of D2-branes in flat space. Our results, however, admit some straightforward generalisations. For example, the ten-dimensional uplift (\ref{G2embeddingGeom}) of the G$_2$-invariant flow depends on the homogeneous nearly-K\"ahler structure of $S^6$ only. It thus remains a valid solution of massive type IIA if $S^6$ is replaced with any nearly-K\"ahler six-manifold. For this reason, as explained in more detail in \cite{Guarino:2015vca}, this particular solution can be uplifted using the universal nearly K\"ahler truncation of \cite{KashaniPoor:2007tr,Cassani:2009ck}. This universality implies that the results of section \ref{subsec:G2flows} also describe the flow triggered by the addition of Chern-Simons terms and G$_2$-invariant matter to the gauge theory defined on a stack of D2-branes probing a G$_2$-holonomy conical singularity. The IR endpoint of the flow is now the Behrndt-Cvetic solution \cite{Behrndt:2004km} constructed out of the nearly-K\"ahler base of the G$_2$-holonomy cone. Similarly, the results of section \ref{subsec:SU3U1flows} apply to describe flows of the gauge theory defined on D2-branes probing a $\textrm{CY}_3 \times \mathbb{C}$ singularity, with $\textrm{CY}_3$ a Calabi-Yau three-fold. In this case, the IR endpoint of the flow corresponds to the generalisation of the $\cN=2$ solution of \cite{Guarino:2015jca} that replaces $\mathbb{CP}^2$ with a suitable K\"ahler-Einstein base. This more general type of $\cN=2$ AdS$_4$ IR phases have also been considered, either analytically or numerically, in \cite{Petrini:2009ur, Lust:2009mb,Tomasiello:2010zz,Fluder:2015eoa}.

While the IR CFT dual to the $\cN=2$ $\textrm{SU}(3) \times \textrm{U}(1)$ AdS$_4$ solution of massive type IIA has been well established and shown to pass non-trivial AdS/CFT tests \cite{Guarino:2015jca}, the same remains to be done for the AdS$_4$ solutions with $\cN=1$ supersymmetry and G$_2$  \cite{Behrndt:2004km} and SU(3) \cite{Varela:2015uca}. It would be very interesting to elucidate these dual CFTs.

\section*{Acknowledgements}

Conversations with Daniel Jafferis, Krzysztof Pilch and Alessandro Tomasiello  are kindly acknowledged. The work of AG is supported by the ERC Advanced Grant no. 246974, {``Supersymmetry: a window to non-perturbative physics''}. JT is supported by  the Advanced ARC project ``Holography, Gauge Theories and Quantum Gravity'' and by the Belgian Fonds National de la Recherche Scientifique FNRS (convention IISN 4.4503.15). OV is supported by the Marie Curie fellowship PIOF-GA-2012-328798.

%%%%%%%%%%%%%%%%%
%%%%%%%%%%%%%%%%%
\appendix
%%%%%%%%%%%%%%%%%
%%%%%%%%%%%%%%%%%

\addtocontents{toc}{\setcounter{tocdepth}{1}}

%%%%%%%%%%%%%%%%%%%%%%%%%%%%%%%
%%%%%%%%%%%%%%%%%%%%%%%%%%%%%%%
\section{Boundary conditions for the numerical integrations}
\label{App:boundary_conds}
%%%%%%%%%%%%%%%%%%%%%%%%%%%%%%%

In this appendix we describe in more detail the boundary conditions necessary to produce the domain wall solutions that we presented in the main text. The integration of the BPS equations (\ref{eq.BPSeqs}) proceeds by shooting from each of the fixed points recorded in table \ref{tab.SU3fps} regarded as an IR fixed point. 

The linearised solution to the BPS flow equations (\ref{eq.BPSeqs}) around the $\cN=1$ G$_2$-point is \eqref{eq.perturbAdS} with coefficients explicitly given by 
\bse\ba \label{BCG2z1}
a_1 & =  0.474 \kappa_ 1 + 0.215 \kappa_ 2 - 0.474 \kappa_ 3 - 0.215 \kappa_ 4 + 
  i (0.215 \kappa_ 1 + 0.098 \kappa_ 2 - 0.215 \kappa_ 3 - 0.098 \kappa_ 4) \ , \\
a_2 & =  0.098 \kappa_ 1 - 0.215 \kappa_ 2 - 0.098 \kappa_ 3 + 
  0.215 \kappa_ 4 + i (-0.215 \kappa_ 1 + 0.474 \kappa_ 2 + 0.215 \kappa_ 3 - 0.474 \kappa_ 4) \ , \\
a_3 & =  0.073 \kappa_ 1 - 0.161 \kappa_ 2 + 0.098 \kappa_ 3 - 0.215 \kappa_ 4 + 
  i (-0.161 \kappa_ 1 + 0.355 \kappa_ 2 - 0.215 \kappa_ 3 - 0.474 \kappa_ 4) \ , \\
a_4 & = 0.355 \kappa_ 1 + 0.161 \kappa_ 2 + 0.474 \kappa_ 3 + 0.215 \kappa_ 4 + 
  i (0.161 \kappa_ 1 + 0.073 \kappa_ 2 + 0.215 \kappa_ 3 + 0.098 \kappa_ 4) \ , 
\end{align}\ese
for $z-z_*$ and
\bse\ba 
\alpha_1 & = -0.355 \kappa_ 1 - 0.161 \kappa_ 2 + 0.355 \kappa_ 3 + 0.161 \kappa_ 4 + 
  i (-0.161 \kappa_ 1 - 0.073 \kappa_ 2 + 0.161 \kappa_ 3 + 
     0.073 \kappa_ 4) \ , \\
\alpha_2 & = -0.073 \kappa_ 1 + 0.161 \kappa_ 2 + 0.073 \kappa_ 3 - 0.161 \kappa_ 4 +
   i (0.161 \kappa_ 1 - 0.355 \kappa_ 2 - 0.161 \kappa_ 3 + 0.355 \kappa_ 4) \ , \\
\alpha_3 & =  0.073 \kappa_ 1 - 0.161 \kappa_ 2 + 0.098 \kappa_ 3 - 0.215 \kappa_ 4 + 
  i (-0.161 \kappa_ 1 + 0.355 \kappa_ 2 - 0.215 \kappa_ 3 + 0.474 \kappa_ 4) \ , \\
\alpha_4 & =  0.355 \kappa_ 1 + 0.161 \kappa_ 2 - 0.161 \kappa_ 3 + 0.355 \kappa_ 4 + 
  i (0.161 \kappa_ 1 + 0.073 \kappa_ 2 + 0.215 \kappa_ 3 + 0.098 \kappa_ 4) \ , \label{BCG2z124}
\end{align}\ese
for $\z-{\z}_*$. Here, $\kappa_1 , \ldots , \kappa_4$ are four independent real integration constants. We provide three decimal places for the numerical coefficients, but we have used larger precision to construct the solutions.

For the $\cN=1$ G$_2$ point to serve as the IR endpoint of a regular domain wall, the integration constants $\kappa_1 , \ldots , \kappa_4$ in (\ref{BCG2z1})--(\ref{BCG2z124}) have to be chosen so that the coefficients $a_i$ and $\alpha_i$ of the exponentials in the linearised solution (\ref{eq.perturbAdS}) with $\tilde \Delta_i>0$ vanish. By inspection of table \ref{tab.deltas}, we see that we need to impose $a_1=a_2=a_3=\alpha_1=\alpha_2=\alpha_3=0$. A solution to these constraints does exist, and this fixes the constants $\kappa_2$, $\kappa_3$ and $\kappa_4$ in terms of $\kappa_1$. This gives
\be
a_4 =\alpha_4 =(1+  0.454 \, i )\kappa_1 \ ,
\ee
and the linearised solution about the IR G$_2$ point thus becomes
\be \label{regG2IRbehaviour}
z-z_*=\z-{\z}_* = (1+  0.454 \, i )\kappa_1 e^{- (1-\sqrt{6})\frac{r}{L_*}} \ .
\ee
The shift $r\to r+\frac{L_*}{1-\sqrt{6}}\log \kappa_1$ of the transverse coordinate is a symmetry of the G$_2$-invariant BPS equations which can be used to set $\kappa_1=1$ without loss of generality. Numerically shooting using the IR boundary condition (\ref{regG2IRbehaviour}), we integrate the BPS equations (\ref{eq.BPSeqs}) into a unique domain wall that approaches the perturbed D2-brane solution (\ref{eq.D2asymptotics}), (\ref{eq.f's_D2}) in the UV, with $c_2 = c_3 = 0$. This flow is plotted in figure \ref{fig.SYMtoG2domainwall}.

Turning now to the $\cN=2$, SU(3)$\times$U(1) fixed point, the linearised solution of the flow equations (\ref{eq.BPSeqs}) about this point is (\ref{eq.perturbAdS}) with
\bse\ba
a_1 & =  0.155 \kappa_ 1 + 0.269 \kappa_ 2 - 0.268 \kappa_ 3 + 
    i (0.269 \kappa_ 1 + 0.466 \kappa_ 2 - 0.464 \kappa_ 3) \ , \\
a_2 & =  0.284 \kappa_ 1 - 0.164 \kappa_ 2 - 0.464 \kappa_ 4 + 
    i (-0.164 \kappa_ 1 + 0.095 \kappa_ 2 + 0.268 \kappa_ 4) \ , \\
a_3 & =  0.466 \kappa_ 1 - 0.269 \kappa_ 2 + 0.464 \kappa_ 4 + 
    i (-0.269 \kappa_ 1 + 0.155 \kappa_ 2 - 0.268 \kappa_ 4) \ , \\
a_4 & = 0.095 \kappa_ 1 + 0.164 \kappa_ 2 + 0.268 \kappa_ 3 + 
    i (0.164 \kappa_ 1 + 0.284 \kappa_ 2 + 0.464 \kappa_ 3) \ , \\
\alpha_1 & = -0.212 \kappa_ 1 -   0.380 \kappa_ 2 + 0.379 \kappa_ 3\ , \\
\alpha_2 & = i (-0.380 \kappa_ 1 + 0.212 \kappa_ 2 + 0.621 \kappa_ 4) \ , \\
\alpha_3 & =  i (0.380 \kappa_ 1 - 0.212 \kappa_ 2 + 0.379 \kappa_ 4)\ , \\
\alpha_4 & =  0.212 \kappa_ 1 + 0.380 \kappa_ 2 + 0.621 \kappa_ 3 \ .
\end{align}\ese
The constants $\kappa_3$ and $\kappa_4$ can be fixed in terms of $\kappa_1$ and $\kappa_2$ to ensure that $a_1=a_2=\alpha_1=\alpha_2=0$, so that the modes with $\tilde \Delta>0$ are turned off for flows that have this point as their IR endpoint. Doing this, we obtain
\bse\ba
a_3 & = \frac{3}{4}\kappa_1- 0.433\kappa_2+i \left(-0.433\kappa_1+\frac{1}{4}\kappa_2 \right) \ , \\
a_4 & = \frac{1}{4}\kappa_1+0.433\kappa_2+i \left(0.433\kappa_1+\frac{3}{4}\kappa_2 \right) \ , \\
\alpha_3 & = i \left( 0.612\kappa_1-0.354 \kappa_2 \right) \ , \\
\alpha_4 & = 0.580 \kappa_1+ 1.004 \kappa_2  \ .
\end{align}\ese
As in the previous case, a shift of the transverse coordinate fixes one of the two constants, say $\kappa_1$, to any positive fixed value. As a result, we end up with a family of flows with $\cN=2$ $\textrm{SU}(3) \times \textrm{U}(1)$ IR endpoint parameterised by $\kappa_2$. 

Some flows in this family have been plotted in figures~\ref{fig.cones} and \ref{fig.AdSdomwalls} with dashed blue lines. They correspond to the values
\be \label{c2ValuesSU3U1}
\kappa_1=10^{-2} \ , \qquad \kappa_2= \left\{ 4.408,\, 1.732, \, 0.981 \right\} \cdot 10^{-2} \ .
\ee
The largest value of $\kappa_2$ in (\ref{c2ValuesSU3U1}) corresponds to the critical domain wall depicted in figure~\ref{fig.AdSdomwalls} whose UV asymptotics is dominated by the G$_2$ fixed point as described in section \ref{CFTtoCFT}. All other values of $\kappa_2$ produce flows with UV asymptotics dominated by the Romans-mass-perturbed D2-brane solution (\ref{eq.D2asymptotics}), (\ref{eq.f's_D2}) with suitable $c_1 , \ldots , c_4$. For example, the smallest value of $\kappa_2$ in (\ref{c2ValuesSU3U1}) produces the dotted blue domain wall in figure~\ref{fig.cones} that runs farthest from the axes. It is shown as a representative of generic behaviour within this family of flows. The central value of $\kappa_2$ in (\ref{c2ValuesSU3U1}) is again special in that it generates the `direct' domain wall that minimises the trajectory between D2-brane UV behaviour and the $\cN=2$ $\textrm{SU}(3) \times \textrm{U}(1)$ IR fixed point. As discussed in the main text, all flows in the family preserve  $\textrm{SU}(3)$ symmetry and $\cN=1$ supersymmetry, except the direct flow for which these are enhanced to $\textrm{SU}(3) \times \textrm{U}(1)$ and $\cN=2$. The `direct' flow corresponds to the central dotted blue line in figure \ref{fig.cones}, and its analytic trajectory is given by equations (\ref{eq:N=2flow}).

Finally, around the fixed point with $\cN=1$ supersymmetry and SU(3) symmetry we observe that the numerical values of $\tilde \Delta_i$ come in two equal pairs. We can thus set $a_{2}=a_{4}=\alpha_{2}=\alpha_{4}=0$ without loss of generality. The remaining coefficients are
\bse\ba
a_1 & =  0.398 \kappa_ 1 - 0.395 \kappa_ 2 - 0.381 \kappa_ 3 + 0.045 \kappa_ 4 + 
  i (-0.395 \kappa_ 1 + 0.602 \kappa_ 2 + 0.045 \kappa_ 3 + 0.381 \kappa_ 4) \ , \\
a_3 & =  0.602 \kappa_ 1 + 0.395 \kappa_ 2 + 0.381 \kappa_ 3 - 0.045 \kappa_ 4 + 
  i (0.395 \kappa_ 1 + 0.398 \kappa_ 2 + 0.045 \kappa_ 3 - 0.381 \kappa_ 4) \ , \\
\alpha_1 & = -0.216 \kappa_ 1 + 0.025 \kappa_ 2 + 0.507 \kappa_ 3 - 0.408 \kappa_ 4 + 
  i (0.025 \kappa_ 1 + 0.216 \kappa_ 2 - 0.408 \kappa_ 3 + 0.492 \kappa_ 4) \ , \\
\alpha_3 & =  0.216 \kappa_ 1 - 0.025 \kappa_ 2 + 0.492 \kappa_ 3 + 0.408 \kappa_ 4 + 
  i (-0.025 \kappa_ 1 - 0.216 \kappa_ 2 + 0.408 \kappa_ 3 + 0.507 \kappa_ 4) \ .
\end{align}\ese
The constants $\kappa_3$ and $\kappa_4$ can be fixed in terms of $\kappa_1$ and $\kappa_2$ to ensure that $a_1=\alpha_1=0$, thus cancelling the mode with $\tilde \Delta>0$. Once this is done we obtain
\bse\ba
a_3 & = \kappa_1+ \kappa_2 \, i \ , \\
\alpha_3 & = 1.150\kappa_1-1.206\kappa_2 + i (0.901 \kappa_1-1.437\kappa_2) \  .
\end{align}\ese
Since both parameters $\kappa_1$ and $\kappa_2$ contribute in this case to the same mode, the sign of $\kappa_1$ or $\kappa_2$ cannot be fixed uniquely. The solutions depicted in figures~\ref{fig.cones} and \ref{fig.AdSdomwalls} were produced with values
\be
\left( \kappa_1 , \kappa_2 \right) =  \left\{ \left( -0.067,-0.1 \right) , \,  \left( 0.114,0.1 \right) , \,  \left( 0.318, 0.1 \right) \right\} \ .
\ee 
The first pair of values corresponds to the critical value where the UV is dominated by the conformal G$_2$ fixed point, depicted by the dashed green curve in Fig.~\ref{fig.AdSdomwalls}. All other values lead to flows with D2-brane UV behaviour. For example, the second pair of values gives the domain wall solution that minimises the trajectory in scalar space and the third one corresponds to a generic flow, the dashed green curve of figure~\ref{fig.cones} that runs the farthest from the axes.

%%%%%%%%%%%%%%%%%
%%%%%%%%%%%%%%%%%
\section{Running of the free energy} \label{App:RunningFE}
%%%%%%%%%%%%%%%%%

The free energy of the different superconformal phases with at least SU(3) flavour symmetry that arise as  IR fixed points of $\cN=8$ SYM upon perturbation with Chern-Simons-matter terms was computed holographically in \cite{Guarino:2015jca,Varela:2015uca}. Here we will extend this computation and determine the running of the free energy  along the entire flows that we constructed in the main text.

The uplift on $S^6$ of the four-dimensional metric $d\tilde{s}_4^2 $ and SU(3)-invariant scalars of the four-dimensional model (\ref{eq.SU3action}) to the ten-dimensional  metric takes the form \cite{Varela:2015uca} 
{\setlength\arraycolsep{2pt}
\begin{eqnarray} \label{KKSU3sectorinIIA}
d\hat{s}_{10}^2 &=&  e^{2A} \,  d\tilde{s}_4^2   \nonumber  \\[5pt]
&&  + g^{-2}   e^{\frac18 (2\phi-\varphi)} X^{1/4}  \Delta_1^{1/2}  \Delta_2^{1/8}  \Big[ \;  e^{-2\phi+\varphi}   X^{-1}  d\alpha^2    \\
&& \qquad  \qquad  \qquad  \qquad   \qquad  \qquad \quad    +    \sin^2 \alpha  \Big( \Delta_1^{-1} ds^2 ( \mathbb{CP}^2 ) + X^{-1} \Delta_2^{-1} (d\psi + \sigma)^2  \Big) \Big] \; . \nonumber
\end{eqnarray}
}Here we have parameterised the four-dimensional scalars as in section 3.1 of \cite{Guarino:2015qaa}. The change of variables into the scalar parameterisation that we have used in the main body of this paper is given by
\begin{equation} \label{VMtrans}
z = \frac{t-i}{t+i} \; , \qquad \textrm{and} \qquad 
\z = \frac{u-i}{u+i} \; , \quad \textrm{with} \; u \equiv -\tfrac12 \sqrt{\zeta^2 + \tilde \zeta^2} + i e^{-\phi} \; .
\end{equation}
The warp factor is given by 
\begin{eqnarray} \label{warping}
e^{2A} = -6\, e^{\frac18 (2\phi-\varphi)} X^{1/4}  \Delta_1^{1/2}  \Delta_2^{1/8}  \, V^{-1} \; ,
\end{eqnarray}
where $V$ is the scalar potential (\ref{eq.SU3Potential}). In (\ref{KKSU3sectorinIIA}), $\alpha$ and $\psi$ are angles on $S^6$ with periods $\pi$ and $2\pi$, respectively, $\sigma$ is a one-form potential for the K\"ahler form on $\mathbb{CP}^2$ and $ ds^2 ( \mathbb{CP}^2 )$ is the Fubini-Study metric normalised so that the Ricci tensor equals six times the metric. Finally,  we have defined
\begin{eqnarray} \label{Delta1}
X & =&  1 + e^{2\varphi} \chi^2 \; , \\[4pt]
\Delta_1 & = & e^{\varphi} \big( 1 + \tfrac14 e^{2\phi}  (\zeta^2 + \tilde \zeta^2)  \big) \sin^2 \alpha  + e^{2\phi-\varphi} \big( 1+ e^{2\varphi} \chi^2 \big) \cos^2 \alpha \; , \\[4pt]
\label{Delta2}
\Delta_2 & = & e^{\varphi}  \sin^2 \alpha  + e^{2\phi-\varphi}   \cos^2 \alpha \; .
\end{eqnarray}
For convenience, we have rescaled the four-dimensional metric $d\tilde{s}_4^2 $ with respect to \cite{Varela:2015uca} by a factor $-6V^{-1}$. This factor evaluates on a critical point of $V$ to the squared AdS radius $L_{*}^2$, see below (\ref{ScaleFactorAdS}). However, the metric $d\tilde{s}_4^2 $ does not need to be the AdS metric corresponding to an IR fixed point, it can rather be any four-dimensional geometry. In fact, here we will be mostly interested in the case in which $d\tilde{s}_4^2 $ is the domain wall metric (\ref{eq.dwmetric}). 

The free energy $F$ is proportional to the inverse of the effective four-dimensional Newton's constant. On the geometry  (\ref{KKSU3sectorinIIA}), (\ref{warping}), this evaluates to
\begin{eqnarray} \label{FreeEnergy1}
F= \frac{16 \pi^3}{(2\pi \ell_s )^8 }  \int_{S^6} e^{8A} \,  \textrm{vol}_6 = -  \frac{96 \pi^3}{(2\pi \ell_s )^8 } \, g^{-6} \, v(S^6) \, V^{-1} \, ,
\end{eqnarray}
where $\ell_s = \sqrt{\alpha^\prime}$ is the string length and $\textrm{vol}_6$ is the volume element corresponding to the metric on the deformed $S^6$ given in (\ref{KKSU3sectorinIIA}), following the conformal factor conventions of \cite{Guarino:2015jca,Varela:2015uca}. The integrand's dependence on the functions $\Delta_1$ and $\Delta_2$ turns out to cancel, leaving solely an $\alpha$ dependence of the form $\sin^5 \alpha$ which integrates into the volume 
\begin{equation}
v(S^6) = \tfrac{16}{15} \, \pi^3
\end{equation}
of the unit radius round six-sphere. Equation (\ref{FreeEnergy1}) already exhibits the expected inverse dependence of the free energy on the scalar potential $V$. However, this expression is written in terms of the classical $D=4$ coupling constants $g$ and $m$ (explicitly and through $V$). Instead, we would like to express the result in terms of the IIA fluxes or, equivalently, the number $N$ of D2-branes and the quantum $k$ of Romans mass. 

In order to do this, note that, at an AdS critical point with at least SU(3) invariance, the cosmological constant scales as $g^2 (m/g)^{-1/3}$ (see table \ref{tab.SU3fps} in the main text). This combination can be taken to set an overall scale not only at a critical point but, in fact, at any point in scalar space. We can thus factorise this dependence from the scalar potential as $V = g^2 (m/g)^{-1/3} \,  \tilde{V}$ and replace $g$ and $m$ by their values in terms of $N$ and $k$ \cite{Varela:2015uca,Guarino:2015jca},
\begin{eqnarray} \label{gmNk}
g^5 = 5 \, v(S^6) \,  (2\pi \ell_s)^{-5} \, N^{-1}  \; , \qquad 
m = k \,  (2\pi \ell_s)^{-1} \; .
\end{eqnarray}
Thus, from (\ref{FreeEnergy1}) we finally obtain
\begin{equation} \label{FreeEnergy2}
F = -96 \cdot 5^{-5/3} \,  \pi^3 \, v(S^6)^{-2/3} \, \tilde{V}^{-1} \, N^{5/3} k^{1/3} \; .
\end{equation} 

At a critical point of the $D=4$ potential with at least SU(3) invariance (see table \ref{tab.SU3fps} in the main text for the supersymmetric points and, more generally, table 3 of \cite{Guarino:2015qaa}), equation (\ref{FreeEnergy2}) produces the free energies given in \cite{Guarino:2015jca,Varela:2015uca}. More generally, (\ref{FreeEnergy2}) is valid at any point of the SU(3)-invariant scalar space, not necessarily at a critical point of the scalar potential. In particular, equation (\ref{FreeEnergy2}) gives holographically the running of the free energy under the renormalisation group flows on the D2-brane with at least SU(3) symmetry that we have considered in this paper. We conjecture that, more generally, equation (\ref{FreeEnergy2}) also holds at any point of the 70-dimensional coset space E$_{7(7)}/\textrm{SU}(8)$ of the full $D=4$ $\cN=8$ dyonically-gauged ISO(7) supergravity, with $\tilde{V}$ given by the full $g=m=1$, $\cN=8$ potential, normalised as in \cite{Guarino:2015qaa}.

%%%%%%%%%%%%%%%%%%%%%%
%%%%%%%%%%%%%%%%%%%%%%
\section{SO(4)-invariant RG flows}
\label{App:SO4}
%%%%%%%%%%%%%%%%%%%%%%

The focus of this paper has been to construct supersymmetric RG flows of the D2-brane field theory triggered by the Romans mass that preserve at least the SU(3) subgroup of the SO(7) R-symmetry of $\cN=8$ SYM in three dimensions. In this appendix, we will briefly touch on similar flows, not necessarily supersymmetric, that have the $\cN=3$ SO(4) critical point \cite{Gallerati:2014xra} of dyonic ISO(7) supergravity as their IR endpoint and preserve at least this SO(4) along the flow. The field theory dual of this critical point was conjectured in \cite{Guarino:2015jca} to correspond to an $\cN=3$ CFT discussed in \cite{Gaiotto:2007qi,Minwalla:2011ma}. The corresponding massive type IIA uplift  has been obtained in \cite{Pang:2015vna} (see also \cite{Pang:2015rwd}) using the consistent truncation formulae of \cite{Guarino:2015jca,Guarino:2015vca}.

\subsection{Flow equations, modes and fixed points}

We will work within the SO(4)-invariant sector of dyonic ISO(7) supergravity described in section 5 of \cite{Guarino:2015qaa}. This sector preserves $\cN=1$ supersymmetry and retains the metric along with four real scalars that parameterise two chiral multiplets. We pack these into two complex fields, $\,\Phi_{1,2}\,$. The bosonic action of this  sector can be written as \cite{Guarino:2015qaa}
\ba
\label{eq.SO4theory}
S & = \frac{1}{16\pi G_4} \int \d^4 x\, \sqrt{-g} \Bigg[ R + \frac{12}{(\Phi_1-\bar{\Phi}_1)^2} \partial_\mu \Phi_1 \partial^\mu \bar{\Phi}_1 + \frac{2}{(\Phi_2-\bar{\Phi}_2)^2} \partial_\mu \Phi_2 \partial^\mu \bar{\Phi}_2  \\
& \qquad \qquad \qquad \qquad \qquad - 8 \left(\frac{2}{3} \left(\Phi_1- \bar{\Phi}_1 \right)^2 \left| \frac{\partial W}{\partial \Phi_1} \right|^2+\frac{1}{4} \left(\Phi_2- \bar{\Phi}_2 \right)^2 \left| \frac{\partial W}{\partial \Phi_1} \right|^2 - 3 |W|^2 \right) \Bigg] \ ,\nonumber
\end{align}
with the scalar potential $V$ canonically expressed in terms of an $\cN=1$ superpotential 
\be
\label{eq.SO4superpotential}
W = g \left( 8 \,\Phi_1^3+6 \,\Phi_1^2 \, \Phi_2 \right) + 2\, m \ 
\ee
(see \cite{Guarino:2015qaa} for further details). This superpotential has a single fixed point. This is also a fixed point of the scalar potential: it is the $\cN=1$, G$_2$ critical point. More generally, the identification $\Phi_1=\Phi_2=-i\,(z+1)/(z-1)$, with $z$ the vector multiplet scalar employed in the main text, reduces the action (\ref{eq.SO4theory}) to that of the G$_2$-invariant sector. 

The scalar potential $V$ displays a number of other extrema which does not share with the superpotential and are therefore not supersymmetric within this truncation. The SO(4)-symmetric AdS point we are interested in is one of these. It is non-supersymmetric within this truncation in spite of being $\cN=3$ within the full $\cN=8$ dyonic ISO(7) supergravity. The reason for this peculiar behaviour was explained in \cite{Guarino:2015qaa}: the three gravitini that remain `massless' at this point transform in a non-trivial representation of SO(4) and are thus projected out from the SO(4)-invariant sector. In the parameterisation that we are using, this SO(4)-invariant point is located at
\be
\label{eq.SO4fixedpoint}
\Phi_{*,1} = 2^{-4/3}\left( -1 + \sqrt{3}\, i \right) c^{1/3} 
\hspace{6mm} , \hspace{6mm} 
 \Phi_{*,2} = 2^{-1/3}\left( 1 + \sqrt{3}\, i \right) c^{1/3} \ ,
\ee
and occurs with the following inverse radius and cosmological constant:
\be
\label{SO4_point}
L_*= \frac{3^{3/4}}{2^{13/6}} \frac{c^{1/6}}{g}
\hspace{6mm} , \hspace{6mm}
V_*= -\frac{32\cdot 2^{1/3}}{\sqrt{3}} \frac{g^2}{c^{1/3}} \ .
\ee
The normalised scalar masses at this point within this sector and the conformal dimensions of the dual operators are summarised in table~\ref{tab.SO4deltas}.

\begin{table}
\centering
\scalebox{0.89}{
\begin{tabular}{@{}cr|cccccc@{}}
\toprule
&& Mode 1 & Mode 2 & Mode 3 & Mode 4  & Relevant oper. & Irrelevant oper. \\ \midrule
\multirow{3}{*}{\makecell{$\cN=3$ \\[1mm] SO(4)}} &
   $M^2L_*^2$    & $3(1-\sqrt{3})$ & $1-\sqrt{3}$ & $1+\sqrt{3}$ & $3(1+\sqrt{3})$   \\[1mm]
&  $\Delta_+$    &  ${\color{blue}\sqrt{3}}$  & ${\color{blue}1+\sqrt{3}}$  & ${\color{blue}2+\sqrt{3}}$  & ${\color{blue}3+\sqrt{3}}$ & 2 & 2 \\
&  $\Delta_-$    & ${\color{blue}3-\sqrt{3}}$ &  ${\color{blue}2-\sqrt{3}}$  & ${\color{red}1-\sqrt{3}}$  &  ${\color{red}-\sqrt{3}}$  \\   \bottomrule
\end{tabular}  
}\normalsize
\caption{Spectrum of SO(4)-invariant scalars around the $\cN=3$ SO(4) fixed point. Blue (red) values correspond to modes compatible with UV (IR) regularity of a domain wall.}
\label{tab.SO4deltas}  
\end{table}

\subsection{SYM to CFT flows}

We now construct an SO(4)-invariant family of domain walls that interpolate between the D2-brane behaviour (\ref{eq.D2_solu}) in the UV and the SO(4)-invariant point (\ref{SO4_point}) in the IR. Since the supersymmetry of the latter is not captured by the model (\ref{eq.SO4theory}), (\ref{eq.SO4superpotential}), we work with the second order Euler-Lagrange equations of motion that derive from it. In particular, the equation of motion of the scale factor $A(r)$ does not decouple from the equations of motion of the scalars $\Phi_i$. It nevertheless still happens to be first order. This allows us to integrate its linearised equation of motion about the IR fixed point as
\be \label{ArSO4}
A(r) = \frac{r}{L_*}\left( 1 +  \frac{c_A}{r} \right) \ ,
\ee
in terms of a unique integration constant, $c_A$, which must be small for the linearised approximation to hold. The linearised equations of motion of the scalars can in turn be integrated about the IR fixed point as
\be \label{PhirSO4}
\Phi_i-\Phi_{*,i} =  \sum_{j=1}^4 \left( b_{+,j,i}\, e^{- \Delta_{+,j} \frac{r}{L_*}}+b_{-,j,i}\, e^{- \Delta_{-,j} \frac{r}{L_*}} \right) \ ,
\ee
in terms of sixteen complex constants $b_{\pm,j,i}$ that depend solely on eight independent real integration constants, respectively associated to each of the conformal dimensions $\Delta_{\pm,j}$, $j = 1 , \ldots , 4$, listed in table~\ref{tab.SO4deltas}.

Next, we proceed with the integration of the entire domain walls. As in the cases covered in the main text, the regularity of an incoming domain wall at the IR fixed point (\ref{eq.SO4fixedpoint}) can be enforced by appropriately choosing boundary conditions. Namely, by imposing relations among the eight integration constants in order to set $b_{\pm,j,i} =0 $ whenever $\Delta_{\pm,j} >0 $. From table \ref{tab.SO4deltas} we see that the only negative exponents  $\,\Delta_{\pm,j}\,$ that can drive a regular domain wall into the IR fixed point via (\ref{PhirSO4}) are the non-normalisable $\Delta_{-,3}=1-\sqrt{3}$ and $\Delta_{-,4}=-\sqrt{3}$. The regularity requirement leaves only two such real integration constants, one of which can be fixed by a shift of the transverse coordinate as in appendix~\ref{App:boundary_conds}. Also, the constant $c_A$ in (\ref{ArSO4}) can  be set to zero without loss of generality, since it just corresponds to a renormalisation of the Minkowski directions in the IR. In conclusion, we find a one-parameter family of SO(4)-invariant domain wall solutions to the Euler-Lagrange equations derived from (\ref{eq.SO4theory})-(\ref{eq.SO4superpotential}) which are smooth in the IR SO(4)-invariant fixed point (\ref{eq.SO4fixedpoint}). Numerical integrations show that the UV of this family of domain walls is dominated by the D2-brane geometry (\ref{eq.D2_solu}).

\begin{figure}
\begin{center}
\includegraphics[width=0.4\textwidth]{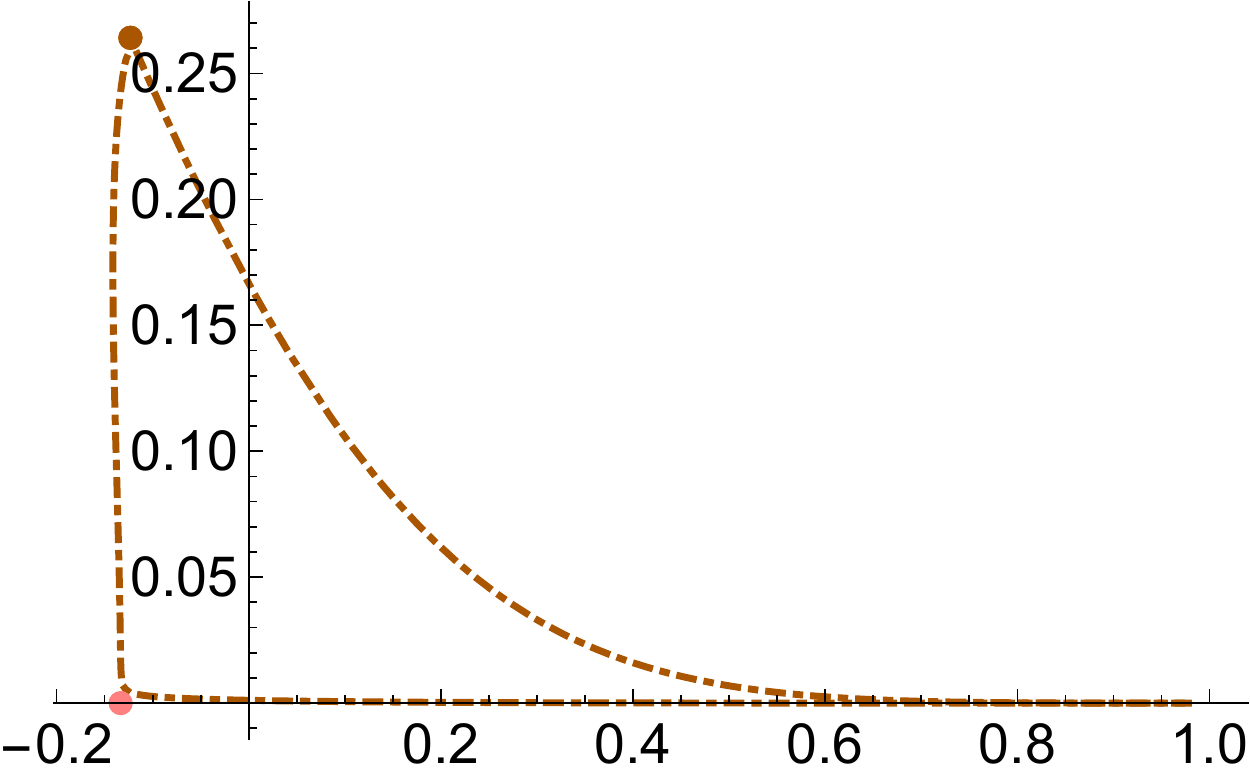}\hspace{1cm}
\hspace{4mm}
\includegraphics[width=0.4\textwidth]{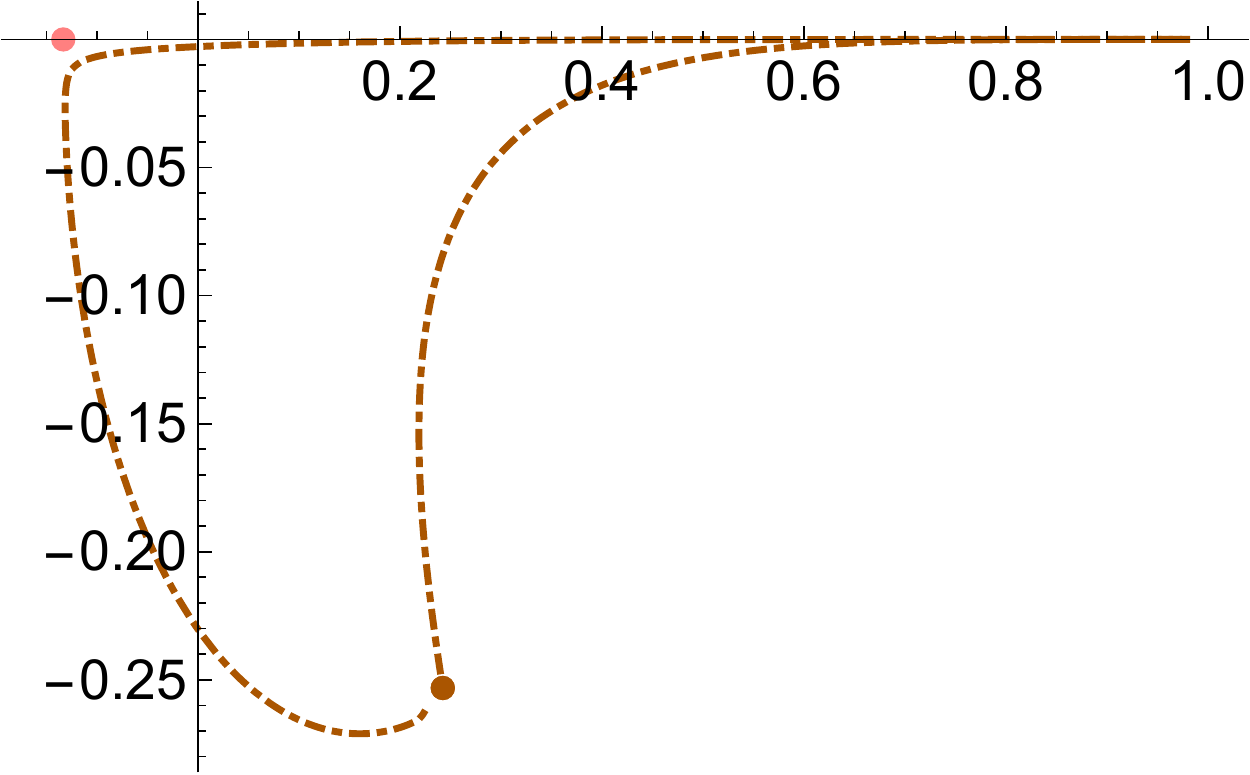}
\put(-200,50){\large Im$z_2$}
\put(-24,110){\large Re$z_2$}
\put(-356,110){\large Im$z_1$}
\put(-241,18){\large Re$z_1$}
\put(-400,110){\small SO(4)}
\put(-100,8){\small SO(4)}
\put(-405,18){\small SO(7)}
\put(-180,110){\small SO(7)}
\caption{Trajectory of a SYM to CFT domain wall solution with the SO(4) fixed point (orange dot) at the IR endpoint.  The non-supersymmetric SO(7) fixed point is depicted with a pink dot.}
\label{fig.SO4susy}
\vspace{-6mm}
\end{center}\end{figure}
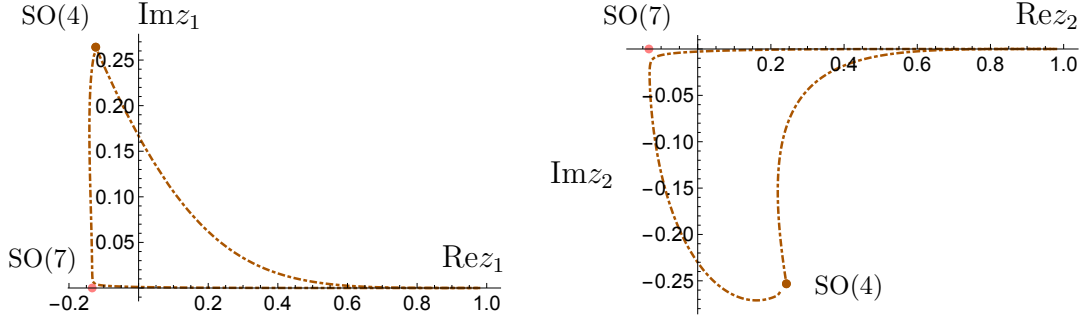

Since we work with second-order equations of motion in this appendix, the flows in this SO(4)-invariant family will typically be non-supersymmetric, even within the full $\cN=8$ dyonic ISO(7) supergravity. The analysis for the construction of the domain wall solutions is, however, very similar to the supersymmetric cases considered in the main text. In figure~\ref{fig.SO4susy} we present two trajectories of domain wall solutions of the Euler-Lagrange equations  in a Poincar\'e disk parameterisation for the scalars
\be
z_j = \frac{\Phi_j-i}{\Phi_j+i} \ .
\ee
The right-most one has only the mode $\Delta_{-,4}$ turned on. The second trajectory has both $\Delta_{-,3}$ and $\Delta_{-4}$, tuned so that this flow experiences walking behaviour dominated by the (unstable) SO(7) point  \cite{DallAgata:2011aa} of dyonic ISO(7) supergravity.

It would be interesting to explicitly check for supersymmetric flows within this family and, more generally, to construct those systematically.

\subsection{Comments on CFT to CFT flows}

From table \ref{tab.SU3fps} and equation (\ref{SO4_point}), the following hierarchy of cosmological constants among the known supersymmetric critical points of dyonic ISO(7) supergravity can be seen to hold:
\be
\label{eq.potentialhierarchyAll}
0 \, > \, V_{*}^{\textrm{G}_2} \,>\, V_{*}^{\textrm{SU}(3)\times \textrm{U}(1)} \,>\, V_{*}^{\textrm{SO}(4)} \,>\, V_{*}^{\textrm{SU}(3)} \ .
\ee
It is then natural to ask whether supersymmetric RG flows from the G$_{2}$ or the SU(3)$\times$U(1) points in the UV to the SO(4) point in the IR exist. The latter flow has been conjectured to exist in \cite{Guarino:2015jca}, following \cite{Gaiotto:2007qi,Minwalla:2011ma}. Similarly, one can ask whether supersymmetric flows between the SO(4) point in the UV and the SU(3) point in the IR exist.

In order to look for such flows, one needs to consider either the full  ISO(7) theory or find a subsector that retains all of these points. A candidate that fulfils the latter requirement is the $\mathbb{Z}_{2} \times \textrm{SO}(3)$-invariant sector discussed  appendix~A of \cite{Guarino:2015qaa}. This is an $\cN=1$ truncation that retains six real scalars, which can be complexified into three chiral fields $\Phi_{1,2,3}$. The SU(3)-invariant sector is recovered upon identifying $\Phi_{2}=\Phi_{3}$ whereas the SO(4)-invariant sector is obtained by setting ${\Phi_{1}=\Phi_{3}}$. Unfortunately, the $\cN=3$ SO(4) point suffers the handicap of being non-supersymmetric also within this sector, for similar reasons as in the SO(4)-invariant sector discussed in the previous subsection. Thus, this subsector does not appear to be suitable to study BPS domain walls among the known supersymmetric extrema of dyonic ISO(7) supergravity.
 
Extending the analysis to include non-supersymmetric domain walls, as in the previous subsection, does not produce flows in the $\mathbb{Z}_{2} \times \textrm{SO}(3)$-invariant sector that were not already contained in the SO(4) sector, at least when the SO(4) point lies at the IR. The reason for this is that the extra two real modes contained in this sector compared to the SO(4) sector cannot trigger new flows. Let us focus on the $\textrm{G}_{2}$, SU(3)$\times$U(1) and SO(4) fixed points. The mass spectra about these points within the $\mathbb{Z}_{2} \times \textrm{SO}(3)$-invariant sector are
\begin{equation}
\begin{array}{llll}
M^2 L_{*}^2 \,\,|_{\textrm{G}_{2}} &=& (2 \times) \, -\frac{1}{6}(11\pm \sqrt{6}) \,\, , \,\, 4\pm\sqrt{6} & ,\\[2mm]
M^2 L_{*}^2 \,\,|_{\textrm{SU(3)$\times$U(1)}} &=& 3\pm\sqrt{17} \,\, , \,\,  (2 \times) \, 2  \,\, , \,\, -\frac{20}{9} \,\, , \,\, -\frac{14}{9} & ,\\[2mm]
M^2 L_{*}^2 \,\,|_{\textrm{SO(4)}}  &=& 3(1-\sqrt{3}) \,\, , \,\,  1-\sqrt{3} \,\, , \,\, 1+\sqrt{3} \,\, , \,\, 3(1+\sqrt{3}) \,\, , \,\, (2 \times) \,  -2 & .
\end{array}
\end{equation}
In addition to the masses in tables~\ref{tab.deltas} and \ref{tab.SO4deltas}, there are two new ones, $-\frac{1}{6}(11\pm \sqrt{6})$, for the G$_{2}$ point and $-\frac{20}{9}$, $-\frac{14}{9}$ for the SU(3)$\times$U(1) point, and a new degenerate one, $\,-2\,$, for the SO(4) point. The latter corresponds to a relevant operator at the SO(4) point, which makes it unsuitable to drive flows into this point when it serves as an IR endpoint.

Let us look at the complete spectrum of the $\,\textrm{SO}(4)=\textrm{SO}(3)_{\textrm{diag}} \times \textrm{SO}(3)_{\textrm{right}}\,$ point within the full ISO(7) theory, in order to figure out a truncation that stands a chance of capturing supersymmetric CFT to CFT flows with the SO(4) point as the IR endpoint. The scalar mass spectrum is given by \cite{Gallerati:2014xra}
\begin{equation}
\label{FullSO4_masses}
\begin{array}{llll}
M^2 L_{*}^2 \,\,|_{\textrm{SO(4)}}  &=& 3(1 - \sqrt{3})^{(\textbf{1},\textbf{1})}   \ , \;   (1 - \sqrt{3})^{(\textbf{5},\textbf{1}) + (\textbf{1},\textbf{1})}  \ , \;   (1 + \sqrt{3})^{(\textbf{5},\textbf{1}) + (\textbf{1},\textbf{1})}   \, ,\,   3(1 + \sqrt{3})^{(\textbf{1},\textbf{1})}  & ; \\[2mm]
& &-2^{(\textbf{3},\textbf{3})+(\textbf{3},\textbf{3})}   \ ; \;   -\tfrac{9}{4}^{(\textbf{2},\textbf{2})}  \ , \; -\tfrac{5}{4}^{(\textbf{2},\textbf{2})+(\textbf{4},\textbf{2})} \ ; \;    0^{(\textbf{2},\textbf{2})+(\textbf{2},\textbf{2})+(\textbf{4},\textbf{2})+(\textbf{3},\textbf{1})+(\textbf{3},\textbf{1})}  & ,
\end{array}
\end{equation}
with dual conformal dimensions 
\begin{equation}
\label{FullSO4_Deltas}
\begin{array}{llll}
\Delta_{+} \,\,|_{\textrm{SO(4)}}  &=&  \sqrt{3}^{\,\,(\textbf{1},\textbf{1})}   \ , \;   (1 + \sqrt{3})^{(\textbf{5},\textbf{1}) + (\textbf{1},\textbf{1})}  \ , \;   (2 + \sqrt{3})^{(\textbf{5},\textbf{1}) + (\textbf{1},\textbf{1})}  \ , \;  (3 + \sqrt{3})^{(\textbf{1},\textbf{1})}  & ; \\[2mm]
& &1^{(\textbf{3},\textbf{3})} \ , \;   2^{(\textbf{3},\textbf{3})}   \ ; \;   \tfrac{3}{2}^{{(\textbf{2},\textbf{2})}} \ , \; \tfrac{5}{2}^{(\textbf{2},\textbf{2})+(\textbf{4},\textbf{2})}  \ ; \;   3^{(\textbf{2},\textbf{2})+(\textbf{2},\textbf{2})+(\textbf{4},\textbf{2})+(\textbf{3},\textbf{1})+(\textbf{3},\textbf{1})}  & .
\end{array}
\end{equation}
Here, the labels $\,^{(\textbf{n},\textbf{m})}\,$ specify $\,\textrm{SO}(4)=\textrm{SO}(3)_{\textrm{diag}} \times \textrm{SO}(3)_{\textrm{right}}\,$ representations. The $\cN=3$ R-symmetry group is identified with $\textrm{SO}(3)_{\textrm{diag}}\,$. The $14$ states in the upper line of (\ref{FullSO4_masses}) (equivalently (\ref{FullSO4_Deltas})) form an $\cN=3$ long gravitino multiplet. The states in the second line correspond to three massless vector multiplets ($3 \times 6$ states in the first block), two semi-short gravitino multiplets ($2 \times 8$ states in the second block) and 22 states (third block) corresponding to the Goldstone bosons of the ISO(7) spontaneous symmetry breaking to SO(4) \cite{Gallerati:2014xra}. 

As already discussed, the desired CFT to CFT flows ending at the SO(4) point can only activate irrelevant modes in the IR ($\Delta_{+}>3$) due to regularity. This selects the scalars with $\,M^2 L_{*}^2\,$ \mbox{normalised} masses and $\,\textrm{SO}(4)=\textrm{SO}(3)_{\textrm{diag}} \times \textrm{SO}(3)_{\textrm{right}}\,$ multiplicities given by  $(1 + \sqrt{3})^{(\textbf{5},\textbf{1}) + (\textbf{1},\textbf{1})}$ and $3(1 + \sqrt{3})^{(\textbf{1},\textbf{1})}$. A truncation keeping these seven fields (among others) is the one retaining the long gravitino multiplet ($14$ scalars) and $6$ out of the $22$ Goldstone bosons. Note that it contains the SO(4)-invariant sector as a subtruncation. Alternatively, it can also be seen as the SO(3)$_{\textrm{right}}$-invariant sector of the ISO(7) theory, which describes an $\cN=4$ supergravity -- the eight gravitini of the full theory decompose as $\textbf{8} \to (\textbf{1}, \textbf{1}) + (\textbf{3}, \textbf{1}) + (\textbf{2}, \textbf{2})$ under $\,\textrm{SO}(4)=\textrm{SO}(3)_{\textrm{diag}} \times \textrm{SO}(3)_{\textrm{right}}\,$ -- coupled to three vector multiplets. The scalar manifold is then identified as $\,\mathcal{M}=\frac{\textrm{SL}(2)}{\textrm{SO}(2)} \times \frac{\textrm{SO}(6,3)}{\textrm{SO}(6)\times \textrm{SO}(3)}$ and accounts for the $\,2+18=14+6\,$ scalars previously discussed. 

We leave the investigation of the possible supersymmetric flows among all superconformal phases of the D2-brane worldvolume field theory, including the $\cN=3$ SO(4) phase, for future work.

\bibliography{references}

\end{document}